\documentclass[conference]{IEEEtran}
\ifCLASSINFOpdf
\else
\fi

\let\labelindent\relax

\usepackage[longend,ruled,vlined]{algorithm2e}
\usepackage{graphicx}
\usepackage{amsmath}
\usepackage{booktabs, subfigure}
\usepackage{color, colortbl}
\usepackage{enumitem}

\newtheorem{definition}{Definition}
\newtheorem{property}[definition]{Property}
\newtheorem{proposition}[definition]{Proposition}
\newtheorem{lemma}[definition]{Lemma}
\newtheorem{theorem}[definition]{Theorem}
\newtheorem{corollary}[definition]{Corollary}

\renewcommand {\marginpar}[1]{}

\newcommand {\rfig}[1]{Figure \ref{fig:#1}}

\newcommand {\bsec}[2]{\section{#1}
                       \label{sec:#2} }

\newcommand {\rsec}[1]{Section \ref{sec:#1}}

\newcommand {\bsubsec}[2]{
                       \subsection{#1}
                       \label{sec:#2} }

\newcommand {\beq}[1]{
                      \begin{equation}
                      \label{eq:#1} }

\newcommand {\beqno}[1]{
                      \begin{eqnarray}
                      \nonumber}

\newcommand {\eeq}       {\end{equation}}
\newcommand {\eeqno}       { && \end{eqnarray}}
\newcommand {\req}[1]{(\ref{eq:#1})}

\newcommand {\bear}[1]{
                       \begin{eqnarray}
                       \label{eq:#1} }

\newcommand {\bearno}[1]{
                       \begin{eqnarray}
                       \nonumber}

\newcommand {\eear}{\end{eqnarray}}
\newcommand {\eearno}{\end{eqnarray}}

\newcommand {\btab}[1]{
                       \begin{table}
                       \centering
                       \begin{tabular}{#1}}
\newcommand {\etab}[3] {
                       \end{tabular}
                       \caption[#3]{#2}
                       \label{tab:#1}
                       \end{table}
                       \vspace{.1in}}
\newcommand {\rtab}[1]{Table \ref{tab:#1}}

\newcommand {\btabular}[1]{\begin{center}
                       \begin{tabular}{#1}}
\newcommand {\etabular}{\end{tabular}
                       \end{center}}

\newcommand {\bdefin}[1]{\begin{definition}\label{def:#1}}
\newcommand {\edefin}       {\end{definition}}

\newcommand {\bpro}[1]{\begin{property}
                      \label{pro:#1} }
\newcommand {\epro}   {\end{property}}

\newcommand {\bprop}[1]{\begin{proposition}
                      \label{prop:#1} }
\newcommand {\eprop}       {\end{proposition}}

\newcommand {\blem}[1]{\begin{lemma}
                      \label{lem:#1}}
\newcommand {\elem}   {\end{lemma}}

\newcommand {\bthe}[1]{\begin{theorem}
                      \label{the:#1} }
\newcommand {\ethe}   {\end{theorem}}

\newcommand {\bcor}[1]{\begin{corollary}
                      \label{cor:#1} }
\newcommand {\ecor}   {\end{corollary}}

\newcommand {\ralg}[1]{Algorithm \ref{alg:#1}}

\newcommand{\hide}[1]{}

\begin{document}
\title{Flow-based Influence Graph Visual Summarization}


\author{\IEEEauthorblockN{Lei Shi}
\IEEEauthorblockA{SKLCS, Institute of Software\\
Chinese Academy of Sciences\\
Beijing 100190, China\\
shil@ios.ac.cn}
\and
\IEEEauthorblockN{Hanghang Tong}
\IEEEauthorblockA{Computer Science\\
City College, CUNY\\
New York, USA\\
tong@cs.ccny.cuny.edu}
\and
\IEEEauthorblockN{Jie Tang and Chuang Lin}
\IEEEauthorblockA{Computer Science\\
Tsinghua University\\
Beijing 100084, China\\
\{jietang, chlin\}@tsinghua.edu.cn}}


%Telephone: (800) 555--1212, Fax: (888) 555--1212}


\maketitle

\begin{abstract}
%\boldmath
Visually mining a large influence graph is appealing yet challenging. People are amazed by pictures of newscasting graph on Twitter, engaged by hidden citation networks in academics, nevertheless often troubled by the unpleasant readability of the underlying visualization. Existing summarization methods enhance the graph visualization with blocked views, but have adverse effect on the latent influence structure. How can we visually summarize a large graph to maximize influence flows? In particular, how can we illustrate the impact of an individual node through the summarization? Can we maintain the appealing graph metaphor while preserving both the overall influence pattern and fine readability?

To answer these questions, we first formally define the influence graph summarization problem. Second, we propose an end-to-end framework to solve the new problem. Our method can not only highlight the flow-based influence patterns in the visual summarization, but also inherently support rich graph attributes. Last, we present a theoretic analysis and report our experiment results. Both evidences demonstrate that our framework can effectively approximate the proposed influence graph summarization objective while outperforming previous methods in a typical scenario of visually mining academic citation networks.

\end{abstract}


% \fi
\IEEEpeerreviewmaketitle

\begin{figure*}[t]
\centering
\includegraphics[width=4.4 in]{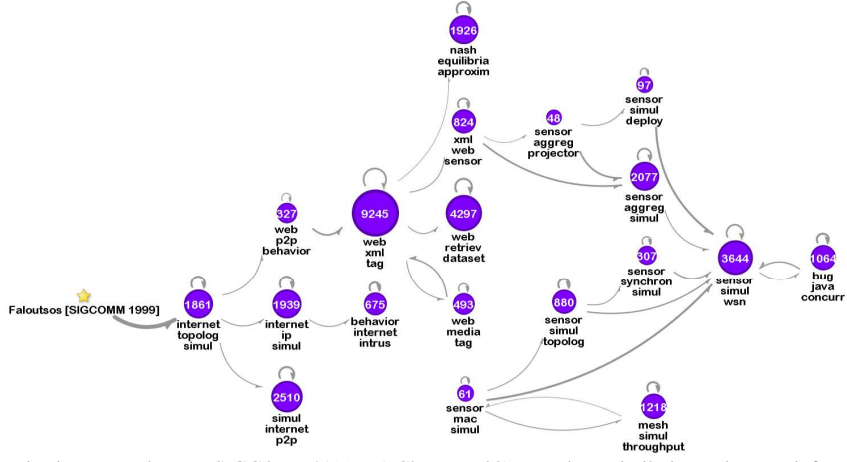}
\vspace{-0.15 in}
\caption{Influence graph summarization on [Faloutsos SIGCOMM'1999] (\#Cluster = 20). Topology similarity and venue information are integrated. Node label gives the cluster size and summary on paper title+abstract normalized by keyword frequency. Link thickness indicates the normalized flow rate.} \label{fig:SIGCOMMPaperExample_20}
\vspace{-0.12 in}
\end{figure*}

\bsec{Introduction}{Intro}

Graphs are prevalent and have become a prevalent platform for the masses to interact and disseminate a variety of information (e.g., influence, memes, opinions, rumors, etc.). {\em How to make sense of an individual's influence in the context of such graphs?} This, which is referred as \underline{I}nfluence \underline{G}raph \underline{S}ummarization (IGS) problem, is the central problem we aim to address in this paper. For example, how does a highly-cited paper impact the research community to raise several topic threads; and consequentially, how do these topics interact with each other and lead to a new multi-disciplinary research direction? How does a senior researcher contribute to multiple research areas by influencing others?

Although closely related, IGS problem bears some subtle difference from the existing work. We briefly review three most relevant topics. First ({\em influence maximization}), in the past decades, many elegant algorithms have been proposed for the so-called influence maximization problem~\cite{Kempe:03}. While effective in identifying {\em who} are most influential in the graph, the question of {\em what makes them influential} largely remains open. Second ({\em graph visualization}), many elaborate layout algorithms have been designed and widely applied in recent years. They can draw medium-sized graphs aesthetically and faithfully, but can not avoid the huge visual clutter on large influence graphs. Third ({\em graph summarization}), many interesting work has been done in the context of graph clustering and compression. These works typically look for coherent/homogeneous regions in graphs by optimizing a pre-defined loss function (e.g., minimizing the inter-cluster connection, maximizing the intra-cluster density, minimizing the total description cost, etc). Despite their own success, most, if not all, of the existing work on graph summarization tends to ignore the specific characteristics of influence graphs and how the end user would visually perceive/read/consume the summarization results.
%Background of the Influence Graph and the problem
%Explanation and expansion of the problem

To be specific, we outline the following design objectives that differentiate our IGS problem from existing works.
\begin{itemize}[leftmargin=.2in]
 \item {\em D1. Flow Rate Maximization}. Quite different from extracting dense clusters on graph, the goal of IGS is to highlight the flow of influence not only within but also across clusters. By maximizing the overall flow rate, IGS-based summarization outlines the strongest interaction among groups of nodes on a graph. For example, \rfig{SIGCOMMPaperExample_20} depicts the influence of the famous power-law paper presented at SIGCOMM'99. The evolution of research topics is revealed, rather than the hot topics themselves.
     %In this sense, it is similar to MDL representation and blockmodeling which consider all blocks between clusters. However, as shown in the objective function defined later, the IGS problem targets a maximization of flow rates, rather than fitting the flows for a minimal cost or homogeneity.
 \item {\em D2. Localized Visualization}. While a large graph can span millions of nodes and prohibit any readable visual summarization, in IGS objective, we switch to summarize the influence of a single node on the graph (called the source node). This localized visualization problem is at least as important as the overall summarization problem. Consider a user navigating the citation graph of computer science papers, after an overview of the entire field, likely she will drill down to a few interested papers and examine their influence separately. %This is known to be the fisheye distortion technique \cite{} or focus+context viewing \cite{} popular in the visualization literature.
 \item {\em D3. Rich Information}. Most influence graphs have rich attributes (e.g., the topic, venue of a scientific paper) and often evolve over time (e.g., the publication date). Incorporating these attributes to enhance the IGS performance poses additional challenges to our work.
\end{itemize}
%Solving the IGS problem is nontrivial due to its distinctive requirements mentioned above. Moreover,

In this paper, we propose a unified framework to generate {\em flow-based}, {\em localized visual} summarization over large-scale influence graphs. The framework provides a seamless, end-to-end pipeline to solve the IGS problem by decomposing it into several key building blocks. It is flexible and admits many existing graph mining algorithms for each of its building blocks. Meanwhile, theoretic analysis shows that our method is equivalent to the kernel k-mean clustering with a carefully designed kernel matrix so that the intra-cluster consistency is also preserved. Finally, we conduct extensive empirical evaluations to validate the effectiveness of the framework. The main contributions of the paper can be summarized as:

\begin{itemize}[leftmargin=.2in]
\item {\em Problem Definition}, to fulfill the design objectives listed above for flow-based visual summarization of large influence graphs (\rsec{Problem});
\item {\em A Unified Framework and Implementation Details}, to solve the IGS problem (\rsec{Frame} and \rsec{Alg});
\item {\em Theoretic Analysis}, to reveal the intrinsic relationship between IGS problem and the existing work (\rsec{Analysis});
\item {\em Comprehensive Evaluation}, to demonstrate the effectiveness and efficiency of the proposed framework (\rsec{Eva}).
\end{itemize}

\bsec{Problem Definition}{Problem}

%Denotation (with table)

\begin{table}
\centering
\caption{Notations.}
\label{tab:notation}
\begin{tabular}
{|l | l | } \hline
SYMBOL & DESCRIPTION \\ \hline \hline
$I$ & influence graph as input\\
$f$ & source node selected by user or algorithm\\ \hline

$G$ & maximal influence graph of $f$ in $I$\\
$v_i$, $N(i)$, $n$ & nodes, neighbor set and \# of nodes in $G$\\
$A$, $a_{ij}$ & adjacency matrix of $G$ and its entries\\
$M^G$,$M^D$,$M^T$& similarity, attribute and time matrix of $G$\\ \hline

$S$ & graph summarization of $G$\\
$\pi_{c}$, $|\pi_{c}|$, $k$ & clusters, cluster size and \# of clusters in $S$\\
$\xi_{s}$, $r(\xi_{s})$, $l$ & flows, flow rate and \# of flows in $S$\\
$\pi_{c(s)}$, $\pi_{d(s)}$ & the source and target cluster of flow $\xi_{s}$\\ \hline
\end{tabular}
\vspace{-0.2 in}
\end{table}

\rtab{notation} lists the notations used throughout the paper. The raw inputs are the influence graph $I$ and the source node $f$ either selected by the user or detected by any existing influence maximization algorithm. Without loss of generality, it is enough to consider a maximal influence graph $G$ of $f$ which is an induced subgraph of $I$ containing all the nodes reachable from $f$ in $I$ (including $f$). Though it is easy to extend the definition to a maximal origin graph by reversing all the links in $I$ or using an union of the two definitions, for relevancy to the IGS problem we stick to the maximal influence graph definition in this paper. Let $G$ have $n$ nodes, denoted as $\{v_i\}_{i=1}^n$. $G$ is represented by the adjacency matrix $A = \{a_{ij}\}_{i,j=1}^{n}$ in which $a_{ij}$ denotes the link weight. $a_{ij} > 0$ if there is a link from $v_i$ to $v_j$.

\emph{Definition 1:} The \textbf{graph summarization} of $G$, denoted as $S$, is a super node-link graph of $G$. The node set of $S$ contains $k$ disjoint and exhaustive node clusters of $G$, denoted as $\{\pi_{c}\}_{c=1}^{k}$ where $|\pi_{c}|$ indicates the number of nodes in cluster $\pi_{c}$. The link set of $S$ contains $l$ flows between the nodes in $S$ (i.e., clusters in $G$), denoted as $\{\xi_{s}\}_{s=1}^{l}$. Each flow $\xi_{s}$ represents the collection of all the links in $G$ from nodes in cluster $\pi_{c(s)}$ to nodes in cluster $\pi_{d(s)}$. The flow rate of $\xi_{s}$ is defined by
\beqno{FlowRate}
r(\xi_{s})=\frac{\sum_{v_i\in\pi_{c(s)},v_j\in\pi_{d(s)}}a_{ij}}{|\pi_{c(s)}||\pi_{d(s)}|}
\eeqno
%\hh{i am not sure what the subscript s in the flow rate mean. can we simply drop s? or denote it as $r(\xi_{c \rightarrow d})$ since the flow rate is defined on a link of $S$}
Note that $S$ can be a partial summarization of $G$, with fewer flows ($l < k^2$) than a full summarization ($l = k^2$). This is desirable for influence graph visualization where huge number of flows and edge crossings can cause unpleasant visual clutter.

%\hh{lei: influence maximization has its special meaning in dm community}
%Objective Function
\emph{Problem 1:} The \textbf{general IGS problem} is defined as finding a graph summarization $S$ with $k$ clusters and $l$ top flows of the maximal influence graph $G$ to optimize the objective function:
\beq{GeneralIGS}
\max~~~\sum_{s=1}^{l}r(\xi_{s})\sqrt{|\pi_{c(s)}||\pi_{d(s)}|}
\eeq

\begin{figure}
\centering
\vspace{-0.1 in}
\includegraphics[width=2.8 in]{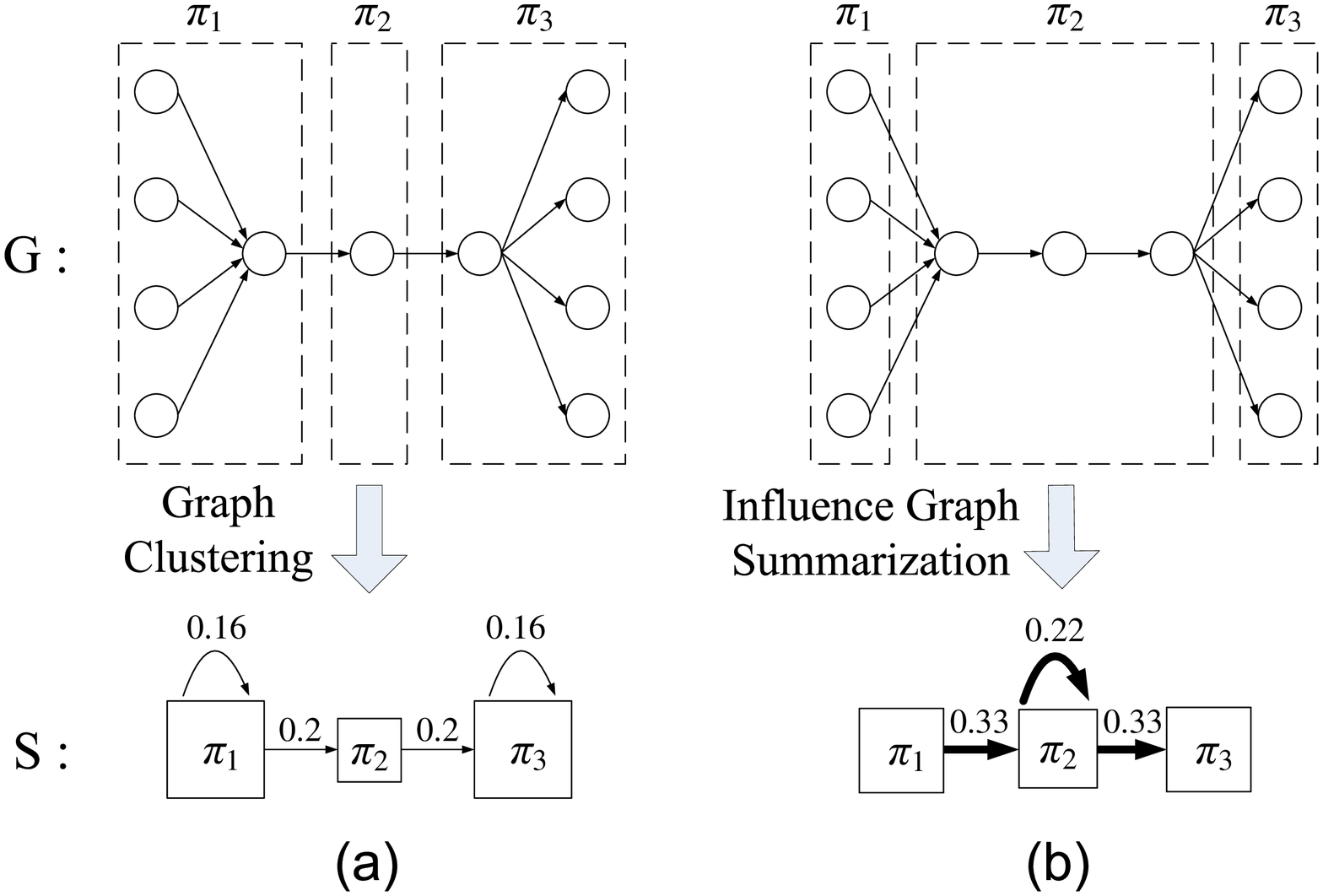}
\vspace{-0.1 in}
\caption{Difference between IGS problem and traditional graph clustering problem. Each dash box in the original graph $G$ becomes a square node in the summarization graph $S$. (a) traditional graph clustering leading to more intra-cluster flows; (b) influence graph summarization exposing denser flows. In $S$, the flow rate is labeled above each link and is mapped to the link thickness visually. We assume a uniform link weight of 1 in the original graph $G$.} \label{fig:ProblemComparison}
\vspace{-0.15 in}
\end{figure}
%Explanation

The general IGS problem defined in \req{GeneralIGS}, although seemingly similar to, is different from the traditional graph clustering problems. Let us explain their difference using the classic ratio association graph clustering problem, whose objective function is shown below.
\beqno{Ratio-Association}
\max~~~\sum_{c=1}^{k}\sum_{i,j\in\pi_{c}}\frac{a_{ij}}{|\pi_{c}|} = \sum_{c=1}^{k}r(\xi_{c})|\pi_{c}|
\eeqno
where $\xi_{c}$ denotes the intra-cluster flow from $\pi_{c}$ to itself.

The IGS objective function is designed to maximize the sum of $l$ selected flows between or within clusters, corresponding to $l$ arbitrary blocks in the adjacency matrix. On the other hand, the ratio association objective maximizes the sum of intra-cluster flows at all the $k$ diagonal matrix blocks. In other words, IGS finds dense flows through summarization which fits well the goal to highlight flows of influence across the graph. This is quite different from the traditional graph clustering objective that finds dense node clusters. An example is given in \rfig{ProblemComparison} for visual comparison.

\begin{figure}[t]
\centering
\includegraphics[width=2.7 in]{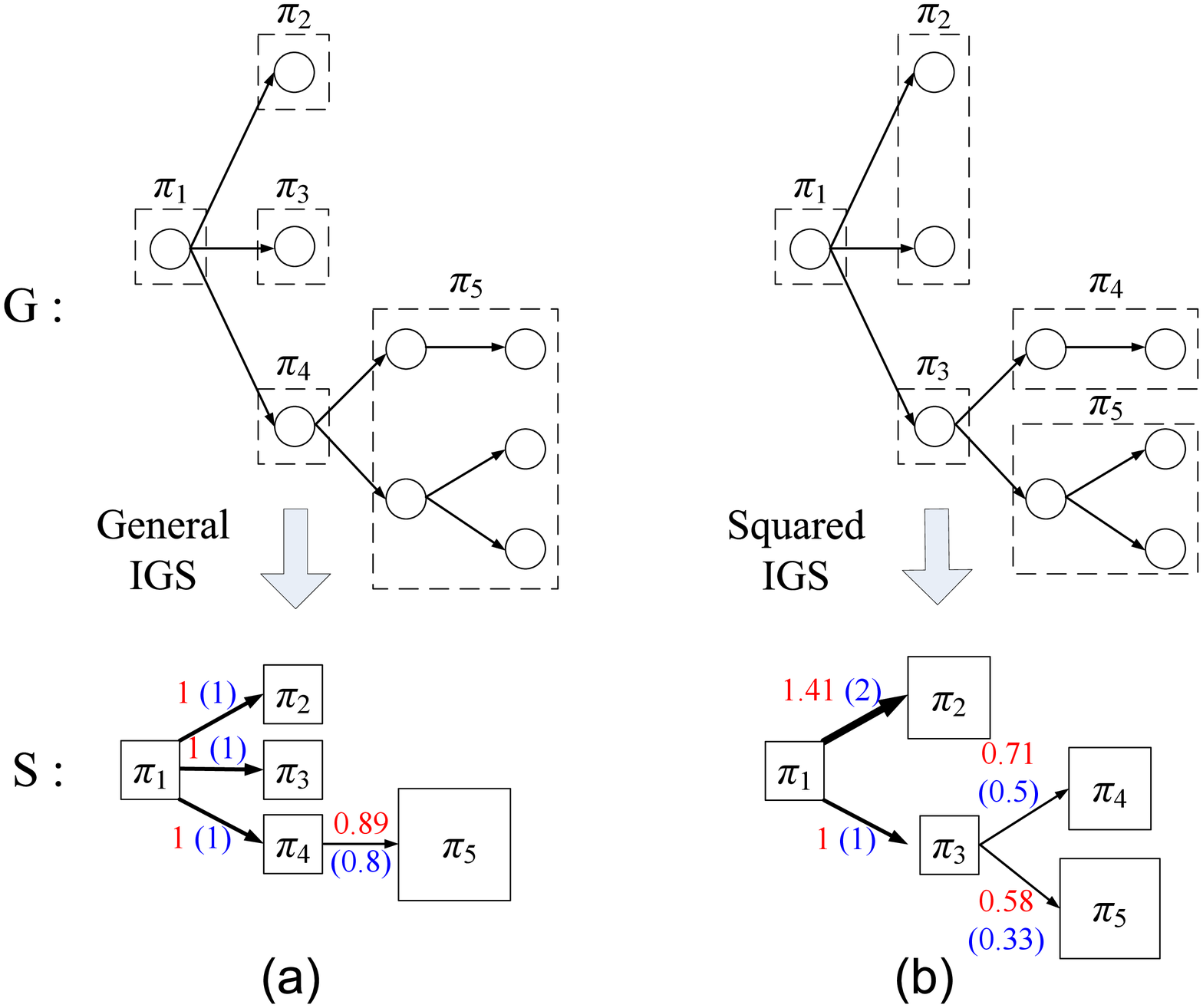}
\vspace{-0.12 in}
\caption{The sample influence graph leading to fragmented flows in the summarization ($k=5$, $l=4$): (a) By the general IGS objective, the resulting graph has two identically-positioned clusters at 1-hop from the source node ($\pi_{2}, \pi_{3}$), the normalized flow rate by \req{GeneralIGS} is labeled in red, favoring this summarization by a sum of $3.89 > 3.70$; (b) Applying the squared IGS objective, the two identical clusters can be merged and more structure of the influence graph is revealed. The squared flow rate by \req{SquaredIGS} is labeled in blue parentheses, having a sum of $3.83 > 3.80$. (best viewed in color)} \label{fig:SquaredComparison}
\vspace{-0.15 in}
\end{figure}

Note that both objective functions are normalized by the square root of the size of clusters/blocks in the adjacency matrix. While this is good for classical graph clustering heuristics, applying the same normalization method on IGS can lead to fragmented flows on the summarization. \rfig{SquaredComparison} illustrates a case with a small influence graph.

%Objective Function
\emph{Problem 2:} The \textbf{squared IGS problem} improves the definition of flow contributions by their squared and normalized flow rate. The new objective function is written as:
\beq{SquaredIGS}
\max~~~\sum_{s=1}^{l}r(\xi_{s})^2|\pi_{c(s)}||\pi_{d(s)}|
\eeq
From the perspective of highlighting influence flows, the squared IGS objective is consistent with the general IGS. Moreover, by applying the square function to the flow rate, it favors large flows more than the general objective. In this sense, heuristically it is better for our influence graph summarization problem with bounded flow number.

\iffalse
\bsubsec{Localization}{Local}

Consider that the maximal influence graph $G$ is derived from the source node $f$, the summarization of $G$ should also be subject to the source node $f$. Here we define a weight matrix $W$ on $G$ localized to the source node $f$. The entries of $W$ is denoted as $\{\omega_{ij}\}_{i,j=1}^n$, indicating the localized weight of each link subject to $f$. As we shall discuss later, $W$ will be constructed to favor the contribution of closer nodes and links to $f$.

\emph{Problem 3:} The \textbf{localized IGS problem} incorporates the weight matrix $W$ to generate a localized graph summarization subject to the source node $f$. The objective function is written as:
\beq{IGS-local}
\max~~~\sum_{s=1}^{l}\omega(\xi_{s})^2|\pi_{c(s)}||\pi_{d(s)}|
\eeq
where $\omega(\xi_{s})$ denotes the localized flow rate of $\xi_{s}$ weighted by the square root of the weight matrix entries to accommodate the squared IGS objective. %\hh{again, check the subscript s}
\beqno{FlowRate-Local}
\omega(\xi_{s}) = \frac{\sum_{v_i\in \pi_{c(s)}, v_j\in \pi_{d(s)}}\sqrt{\omega_{ij}}a_{ij}}{|\pi_{c(s)}||\pi_{d(s)}|}
\eeqno
\fi

\bsec{Framework}{Frame}

In this section, we propose a unified framework to solve the IGS problem, including an end-to-end pipeline, the algorithm to summarize influence structure from graph topology, and the extension to incorporate graph attribute and time information.

\bsubsec{End-to-End Pipeline}{Overview}

\begin{figure}[t]
\centering
\includegraphics[width=3.5 in]{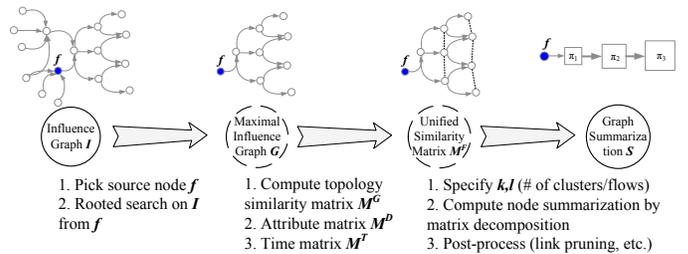}
\vspace{-0.3 in}
\caption{The framework to solve the influence graph summarization problem.}
\label{fig:Framework}
\vspace{-0.17 in}
\end{figure}

%Block diagram
We propose an end-to-end pipeline, shown in \rfig{Framework}. The framework decomposes the IGS problem into several building blocks. Initially, the maximal influence graph $G$ is computed from the input graph $I$ by a breadth-first or depth-first search starting from the source node $f$. Over the maximal influence graph $G$, three processing components work in parallel to generate three matrices on the graph: the topology similarity matrix, and the optional attribute and time matrices. The core of our framework is the decomposition of the topology similarity matrix to generate $k$ node clusters for the summarization. We carefully design the topology similarity matrix to ensure that the graph summarization approximates the flow rate maximization objective (See \rsec{Analysis} for detailed analysis). The attribute and time matrices can be incorporated to augment the similarity matrix before decomposition so as to optimize the summarization towards graph attributes. The requirement of the $l$ flows in the summarization is handled by link pruning using either ranking-based filtering or the maximum spanning tree algorithm. The proposed pipeline is flexible and admits many existing graph mining algorithms for each of its building blocks. On the other hand, by itself, none of these existing algorithms is sufficient to solve the IGS problem.

\bsubsec{Node Summarization}{Desc}

%Symmetric NMF
Node summarization is the key building block of our proposed pipeline. It takes the topology similarity matrix $M^G$ as the input and generates $k$ node clusters. We propose a matrix decomposition based solution. Its rationality as well as the details of the similarity matrix $M^G$ will be discussed in \rsec{Analysis} and \ref{sec:Alg}, respectively.

Over the similarity matrix $M^G$, the decomposition employs a Symmetric version of the Nonnegative Matrix Factorization (SymNMF \cite{SymNMF}) which optimizes:
\beq{SymNMF}
\min_{H\geq0}||M^G-HH^T||_F^2
\eeq
where $||\cdot||_F$ denotes the Frobenius norm of the matrix. $H=\{h_{ij}\}$ is a $n$ by $k$ matrix indicating the cluster membership assignment of nodes in $G$: $v_i$ will be clustered into $\pi_c$ if $h_{ic}$ is the largest entry in the $i$th row of $H$.

%Localized NMF
\iffalse
To solve the localized version of IGS problems, the weight matrix of $G$ (denoted as $W$) is incorporated. The SymNMF decomposition becomes the weighted SymNMF:
\beq{SymNMF_Weight}
\min_{H\geq0}||W\odot(M^G-HH^T)||_F^2
\eeq
where $\odot$ indicates the Hadamard (by element) product of matrices. In our framework, $W$ is computed by random walk with restart (RWR~\cite{rwr}) on $G$ from the source node $f$. The RWR weight is known to well represent the ranking of each node subject to the source node $f$. Formally, the weight matrix $W$ is constructed by
\beqno{WeightMatrix}
\omega_{ij} = \left\{
             \begin{array}{lr}
             \omega_{ii}, & i = j \\
             \sqrt{\omega_{ii}\omega_{jj}}, & i \neq j
             \end{array}
\right.
\eeqno
where the diagonal elements of $W$ are set by the RWR weight of each node from $f$. Under this construction, the objective function in \req{SymNMF_Weight} is converted to that of a standard SymNMF.
\beq{SymNMF_Weight2}
\min_{H\geq0}||W\odot M^G-(W_2\odot H)(W_2\odot H)^T||_F^2
\eeq
where $W_2=\{\sqrt{\omega_{ii}}\}_{i,j=1}^n$. This provides us extra advantage because the algorithm for weighted SymNMF does not converge as well as that of the standard SymNMF.
\fi

%An example

\bsubsec{Generalizations}{Ext}

Our framework can be extended to incorporate node attributes and their time information on graph (e.g., the research field attribute and the publication date of a paper). The extension takes in two separate matrices computed from the influence graph $G$. The attribute augmentation matrix is constructed to reflect the pairwise similarities among graph nodes in terms of the specific attribute. Consider an attribute affinity adjacency matrix $A^D=\{a^D_{ij}\}^n_{i,j=1}$, where $a^D_{ij} > 0$ if $v_i$ and $v_j$ have the same value on a selected node attribute.  Denote the attribute augmentation matrix as $M^D=\{m^D_{ij}\}^n_{i,j=1}$, $M^D$ is computed from $A^D$ by
\beqno{Attribute-Aug}
m^D_{ij} = \left\{
             \begin{array}{lr}
             \Lambda_{aug}, & i = j~~or~~a^D_{ij} > 0\\
             1, & a^D_{ij} = 0
             \end{array}
\right.
\eeqno
where $\Lambda_{aug}$ controls the degree of augmentation. We find $\Lambda_{aug}=2$ is an effective setting in general.

Similarly, we construct the time decaying matrix $M^T=\{m^T_{ij}\}^n_{i,j=1}$ to reflect the similarity among graph nodes on the associated time. For example, the papers in the same year will have a high similarity entry on $M^T$. Denote the time attribute of graph nodes on $G$ as $\{t_i\}^n_{i=1}$ (unit year), the time decaying matrix $M^T$ is computed by
\beqno{Time-Aug}
             m^T_{ij} = {\lambda_{Decay}}^{-|t_i - t_j|}
\eeqno
where $\lambda_{Decay}$ controls the rate of similarity decaying over time. Using the median of cited half-life time of sci-indexed CS journals, we compute a default value of $\lambda_{Decay}=1.11$.

The attribute and time matrices are then used to extend the topology similarity matrix to a unified form.
\beq{Fusing}
M^F=M^G \odot M^D \odot M^T
\eeq
The final SymNMF objective becomes
\beq{Final-SymNMF}
\min_{H\geq0}||M^F-HH^T||_F^2
\eeq

\bsec{Equivalence Analysis}{Analysis}
In this section, we present theoretic analysis, to explain the rationality behind our matrix decomposition based solution. We start with deriving an approximate objective function of the IGS problem. Then we show that such an objective is equivalent to the kernel k-mean clustering by choosing an appropriate kernel matrix. Finally, the kernel k-mean clustering can be solved by SymNMF. %This equivalence explains the methods proposed in \rsec{Frame}.

\bsubsec{Approximation of IGS problem}{Appro}

Consider the objective function in \req{SquaredIGS}, the optimization requires maximizing over two types of variables: $\{\pi_c\}_{c=1}^k$, the node cluster membership assignment; and $\{\xi_s\}_{s=1}^l$, the selected top flows. The simultaneous optimization of these two classes of variables is hard due to the non-linear and combinatorial nature of the problem. Here we consider a two-step approximation that first maximizes the sum of all the flows over the node cluster assignment, then maximizes the sum of the top $l$ flows given the cluster assignment. This is feasible with an appropriate $l$ (e.g. $l=2k$), because the top $l$ flows contribute the most part of the overall flow rate after applying the square function, as shown in \rsec{Eva}. Formally, the approximate objective function becomes:
\beq{square-IGS}
\max~\sum_{s=1}^{k^2}r(\xi_{s})^2|\pi_{c(s)}||\pi_{d(s)}| = \sum_{c,d=1}^{k}\frac{(\sum_{i\in\pi_{c}, j\in\pi_{d}}a_{ij})^2}{|\pi_{c}||\pi_{d}|}
\eeq
\beq{square-IGS-2}
\max~\sum_{s=1}^{l}r(\xi_{s})^2|\pi_{c(s)}||\pi_{d(s)}|~~~~~~~~~~~~~given~\{\pi_c\}_{c=1}^k
\eeq
The second part of the optimization can be solved by selecting $l$ top flows with the largest size-normalized flow rate.

%Kernel K-Mean
\bsubsec{Kernel K-Mean Clustering}{Kernel}

According to \cite{KernelKMean}, the kernel k-mean clustering (KM) is defined as follows. Given $n$ data vectors $\{x_{i}\}_{i=1}^{n}$ with kernel function $\phi(x_{i})$, KM method groups the data vectors into $k$ non-overlapping clusters $\{\pi_{c}\}_{c=1}^{k}$ based on the objective function
\beqno{Kernel-KMean-Pre1}
\min~\sum_{c=1}^{k}\sum_{i\in\pi_{c}}||\phi(x_{i})-m_{c}||^{2}~~where~m_{c}=\frac{\sum_{i\in\pi_{c}}\phi(x_{i})}{|\pi_{c}|}
\eeqno
Expand $||\phi(x_{i})-m_{c}||^{2}$ into
\beqno{middle-1}
\phi(x_{i})\cdot\phi(x_{i}) - \frac{2\sum_{j\in\pi_{c}}\phi(x_{i})\cdot\phi(x_{j})}{|\pi_{c}|} +
\frac{\sum_{j,l\in\pi_{c}}\phi(x_{j})\cdot\phi(x_{l})}{|\pi_{c}|^{2}}
\eeqno
Because
\beqno{middle-2}
\sum_{c=1}^{k}\sum_{i\in\pi_{c}} \frac{\sum_{j\in\pi_{c}}\phi(x_{i})\cdot\phi(x_{j})}{|\pi_{c}|} = \sum_{c=1}^{k}\sum_{i\in\pi_{c}}\frac{\sum_{j,l\in\pi_{c}}\phi(x_{j})\cdot\phi(x_{l})}{|\pi_{c}|^{2}}
\eeqno
The objective function of KM clustering can be written as
\beqno{Kernel-KMean-Pre2}
\min~~~\sum_{c=1}^{k}\sum_{i\in\pi_{c}}[\phi(x_{i})\cdot\phi(x_{i}) - \frac{\sum_{j\in\pi_{c}}\phi(x_{i})\cdot\phi(x_{j})}{|\pi_{c}|}]
\eeqno
As $\sum_{c=1}^{k}\sum_{i\in\pi_{c}}{\phi(x_{i})\cdot\phi(x_{i})}$ is constant, it is equivalent to
\beq{Kernel-KMean-Final}
\max~~~\sum_{c=1}^{k}\sum_{i,j\in\pi_{c}}{\frac{\phi(x_{i})\cdot\phi(x_{j})}{|\pi_{c}|}}
\eeq
Introduce the heuristic of 1-hop bidirectional common neighbor as the similarity measure (CommonNeighbor), we can compute a topology similarity matrix by
\beqno{Kernel-KMean-Matrix}
K = \frac{AA^T + A^TA}{2}~~~where~~k_{ij} = \sum_{t=1}^{n}\frac{a_{it}a_{jt}+a_{ti}a_{tj}}{2}
\eeqno
If we use $K$ as the kernel matrix in KM clustering and substitute $k_{ij}$ for $\phi(x_{i})\cdot\phi(x_{j})$, \req{Kernel-KMean-Final} becomes
\bear{Kernel-KMean-CommonNeighbor}
\max & \sum_{c=1}^{k}\frac{1}{|\pi_{c}|}\sum_{i,j\in\pi_{c}}\sum_{t=1}^{n}\frac{a_{it}a_{jt}+a_{ti}a_{tj}}{2}\nonumber\\
= & \sum_{c=1}^{k}\sum_{t=1}^{n}\sum_{i,j\in\pi_{c}}\frac{a_{it}a_{jt}+a_{ti}a_{tj}}{2|\pi_{c}|} \nonumber\\
= & \sum_{c=1}^{k}\sum_{t=1}^{n}\frac{(\sum_{i\in\pi_{c}}a_{it})^2+(\sum_{i\in\pi_{c}}a_{ti})^2}{2|\pi_{c}|}\nonumber\\
= & \sum_{c=1}^{k}\sum_{j=1}^{n}\frac{(\sum_{i\in\pi_{c}}a_{ij})^2+(\sum_{i\in\pi_{c}}a_{ji})^2}{2|\pi_{c}|}\nonumber\\
= & \sum_{c,d=1}^{k}\sum_{j\in\pi_{d}}\frac{(\sum_{i\in\pi_{c}}a_{ij})^2+(\sum_{i\in\pi_{c}}a_{ji})^2}{2|\pi_{c}|}
\eear

\bsubsec{Equivalence}{Equiv}

%General IGS

\begin{figure}[t]
\centering
\includegraphics[width=3 in]{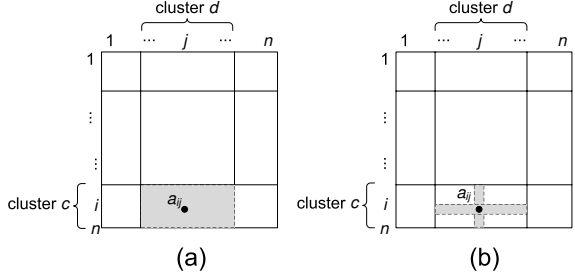}
\vspace{-0.1 in}
\caption{The weighting schema comparison in two objective functions: (a) influence graph summarization using the entire block; (b) kernel k-mean using the block's column and row.} \label{fig:DensityComparison}
\vspace{-0.15 in}
\end{figure}

Let us compare the objective functions in \req{square-IGS} and \req{Kernel-KMean-CommonNeighbor}. They are in similar forms if we re-formulate \req{square-IGS} into
\iffalse
\begin{align}
& \sum_{c,d=1}^{k}\frac{(\sum_{i\in\pi_{c}, j\in\pi_{d}}a_{ij})^2}{|\pi_{c}||\pi_{d}|} = \sum_{c,d=1}^{k}\sum_{i\in\pi_{c}, j\in\pi_{d}}a_{ij}(\frac{\sum_{p\in\pi_{c}, q\in\pi_{d}}a_{pq}}{|\pi_{c}||\pi_{d}|})\nonumber\\
& = \sum_{i,j=1}^{n}a_{ij}w_{ij}^{IGS} \nonumber
\end{align}
\fi
\begin{align}
\sum_{c,d=1}^{k}\sum_{i\in\pi_{c}, j\in\pi_{d}}a_{ij}(\frac{\sum_{p\in\pi_{c}, q\in\pi_{d}}a_{pq}}{|\pi_{c}||\pi_{d}|}) = \sum_{i,j=1}^{n}a_{ij}w_{ij}^{IGS} \nonumber
\end{align}
\beq{IGS-weight}
where ~~~ w_{ij}^{IGS} = \frac{\sum_{p\in\pi_{c},q\in\pi_{d}}a_{pq}}{|\pi_{c}||\pi_{d}|}~(i\in\pi_{c},j\in\pi_{d})
\eeq
and re-formulate \req{Kernel-KMean-CommonNeighbor} into
\iffalse
\begin{align}
&\sum_{c,d=1}^{k}\sum_{j\in\pi_{d}}\frac{(\sum_{i\in\pi_{c}}a_{ij})^2+(\sum_{i\in\pi_{c}}a_{ji})^2}{2|\pi_{c}|}\nonumber\\
&=\sum_{c,d=1}^{k}\frac{1}{2|\pi_{c}|}[\sum_{j\in\pi_{d}}\sum_{i\in\pi_{c}}a_{ij}(\sum_{p\in\pi_{c}}a_{pj})+ \sum_{j\in\pi_{d}}\sum_{i\in\pi_{c}}a_{ji}(\sum_{q\in\pi_{c}}a_{jq})]\nonumber\\
&=\sum_{i,j=1}^{n}a_{ij}w_{ij}^{KM} \nonumber
\end{align}
\fi
\begin{align}
&\sum_{c,d=1}^{k}\frac{1}{2|\pi_{c}|}[\sum_{j\in\pi_{d}}\sum_{i\in\pi_{c}}a_{ij}(\sum_{p\in\pi_{c}}a_{pj})+ \sum_{j\in\pi_{d}}\sum_{i\in\pi_{c}}a_{ji}(\sum_{q\in\pi_{c}}a_{jq})]\nonumber\\
&=\sum_{i,j=1}^{n}a_{ij}w_{ij}^{KM} \nonumber
\end{align}
\beq{Kernel-KMean-weight}
where~~~w_{ij}^{KM}=\frac{\sum_{p\in\pi_{c}}a_{pj}}{2|\pi_{c}|}+\frac{\sum_{q\in\pi_{d}}a_{iq}}{2|\pi_{d}|}~(i\in\pi_{c},j\in\pi_{d})
\eeq
Thus, both IGS and KM aim to maximize the weighted sum of graph adjacency matrix entries. In IGS, the weight of each entry is defined by the density of the belonging matrix block (or flow). In KM, the weight is defined by the average density of the column and row of the belonging matrix block. This is illustrated in \rfig{DensityComparison}. Note that the heuristic of the CommonNeighbor based k-mean clustering is to put the graph nodes with similar in- and out-neighborhoods together. The resulting matrix blocks after the clustering tend to have uniform density distributions inside each block. Therefore, the density of the cross shape area in \rfig{DensityComparison}(b) is a good approximation of the density of the shaded block area in \rfig{DensityComparison}(a), which explains the rationality of using kernel k-mean clustering to the general IGS problem.

Furthermore, it is known that the kernel k-mean clustering problem is equivalent to the trace maximization problem:
\beqno{Trace-Maximization}
\max_{H^TH=I,H\geq0}Tr(H^TKH)
\eeqno
where the kernel matrix $K$ equals the topology similarity matrix $M^G$ computed by CommonNeighbor.
The trace maximization problem can then be solved by SymNMF under spectral relaxations \cite{SymNMF}.

%Weighted IGS

\iffalse
The localized IGS problem can be studied similarly. Compare the objective function in \req{IGS-local} with \req{SquaredIGS}, we can see that the only difference is to replace $a_{ij}$ with $\sqrt{\omega_{ij}}a_{ij}$. After the same analysis as above, the SymNMF to solve becomes:
\beqno{SymNMF_Weight_Analysis}
\min_{H\geq0}||W\odot M^G-HH^T||_F^2
\eeqno
which leads to the solution to \req{SymNMF_Weight2}, because only the relative ranking of entries in each row of $H$ is used for the cluster assignment.
\fi

%Exact method

\iffalse
In fact, an exact equivalence to the IGS objective function in \req{IGS-weight} is possible. The new kernel matrix $K^*$ is devised from the cluster-column-normalized adjacency matrix $A^{*G}=\{a_{ij}^{*G}\}_{i,j=1}^n$ of $G$.
\bear{New_Kernel}
a_{ij}^{*G}=\frac{\sum_{p\in \pi_c}a_{pj}}{|\pi_c|}~~(i\in\pi_{c}) \nonumber\\
K^*=(A^{*})^TA^{*}~~~~~~~~~~~~
\eear
This more accurate kernel matrix depends on the clustering assignment which comes from the optimization result. Later, we will propose an iterative algorithm to achieve that. On the other hand, the mirrored kernel matrix construction with cluster-row-normalized adjacency matrix is possible and will also achieve the exact equivalence to the IGS objective.
\fi

\bsec{Implementation Details}{Alg}

In this section, we provide some additional implementation details. As shown in \rfig{Framework}, our framework involves four kinds of algorithm-driven building blocks. The rooted graph search follows the standard BFS/DFS implementation. %Here only note that the restart probability of RWR is set to 0.8 by default.
Below we describe details for similarity matrix computation, node summarization and the link pruning for post-processing of the summarization.

%Similarity matrix computation (common neighbor, iterative, SimRank)
%\bsubsec{Similarity Computation}{Similarity}

{\bf Similarity Matrix Computation}. In \rsec{Analysis}, we have shown that using the heuristic of common neighbors to construct the similarity matrix (CommonNeighbor) can approximate the objective function of the squared IGS problem. This algorithm runs fast even for very large graphs due to a complexity of $O(md^2)$ where $m$ is the number of links in $G$ and $d$ is the average node degree. We have implemented three versions of the algorithm and it is shown that bidirectional CommonNeighbor is generally better than one-directional forward or backward CommonNeighbor.

%NMF Decomposition (ED for initialization, multiplicative update)
%\bsubsec{Node Clustering}{SymNMF}

{\bf Node Summarization with SymNMF}. The node summarization is done by applying SymNMF on similarity matrix $M^G (M^F)$, and using the factorized matrix $H$ for cluster membership assignment. In our implementation, we apply the iterative SymNMF solver with the multiplicative updating rule in \cite{SymNMF} which guarantees convergence. In this iterative algorithm, the initialization of $H$ is critical to the final result. We introduce nonnegative eigenvalue decomposition similar to the method in \cite{NNDSVD} to compute a good initial factorization.

%Over this initialization, we compute the cluster assignment of the source node $f$ by its largest entry in $H$, denoted as $\pi(f)$. The other entries of $H$ that are related to the source node $f$ and the cluster $\pi(f)$ are cleared to to zero. Due to the nature of multiplicative updating, the cluster assignment of the source node is guaranteed to be unchanged and isolated during the iteration.

%Post-process of the influence graph (threshold and connect approach (consider loop edge), MST approach)
%\bsubsec{Link Pruning}{Pruning}
{\bf Link Pruning.} The graph summarization by SymNMF needs further post-processing to select $l$ top flows for the final summarization $S$. According to \req{square-IGS-2}, the top flows can be extracted after ranking by the normalized flow rate. The other flows are then filtered out. This is illustrated in \ralg{Pruning}. Notice that in the link recovery section of the algorithm, we introduce a constraint to keep a connected graph in the summarization. It is achieved by adding back the most dense flow going to each node cluster. An alternative choice is to use the maximum spanning tree (MST) algorithm \cite{MST}.

\begin{algorithm}[t]
 \label{alg:Pruning}
 \SetKwFunction{FnRankFilter}{\bf RankFilter}
\SetKwInOut{Input}{Input}\SetKwInOut{Output}{Output}
 \Input{Initial summarization $S_0\sim\{V,E\}$, \# of flows $l$, $V=\{\pi_i\}_{i=1}^k$, $E=\{\xi_s\}_{s=1}^{k^2}$, flow rate $r(\xi_s)$}
 \Output{Final summarization $S$}

\FnRankFilter{$S_0$}:\\
 \Begin{
 $S \leftarrow S_0$\;
 \For(\tcp*[f]{rate normalization}){$s \leftarrow 1$ \KwTo{$k^2$}}{
     $r(\xi_s) \leftarrow r(\xi_s)\sqrt{|\pi_c||\pi_d|}$, $\xi_s\sim(\pi_c, \pi_d)$\;
 }
 sort $E$ by $r(E)$ in decreasing order\;
 \For(\tcp*[f]{pruning}){$s \leftarrow l+1$ \KwTo{$k^2$}}{
     remove $E(s)$ from $S$\;
 }
  \For(\tcp*[f]{link recovery}){$i \leftarrow 1$ \KwTo{k}}{
     $E_i \leftarrow $  subset of $E$ having $\pi_i$ as destination\;
     sort $E_i$ by $r(E_i)$ in decreasing order\;
     \If{$E_i(0) \not\in S$}{
      add $E_i(0)$ to $S$\;
     }
 }
}

\caption{Link Pruning Algorithm.}
\end{algorithm}

%Visual exploration (RRE, hierarchical computation)
%\bsubsec{Visual Exploration}{Exploration}

We implement the proposed framework and algorithms in Java, which provides excellent UI library for visualization. The main computation routines are built on ParallelColt package \cite{ParallelColt} to optimize for multi-threading and sparse matrix operations. The speed of some core matrix decompositions (e.g., Eigenvalue) are further improved by invoking ARPACK (for sparse matrix) and LAPACK (for dense matrix) implementation \cite{IntelMKL} through JNI invocations.

%objective function performance
\begin{figure*}[t]
\centering
\subfigure[$k$ = 10, $l$ = 10]{\includegraphics[height=1.8 in]{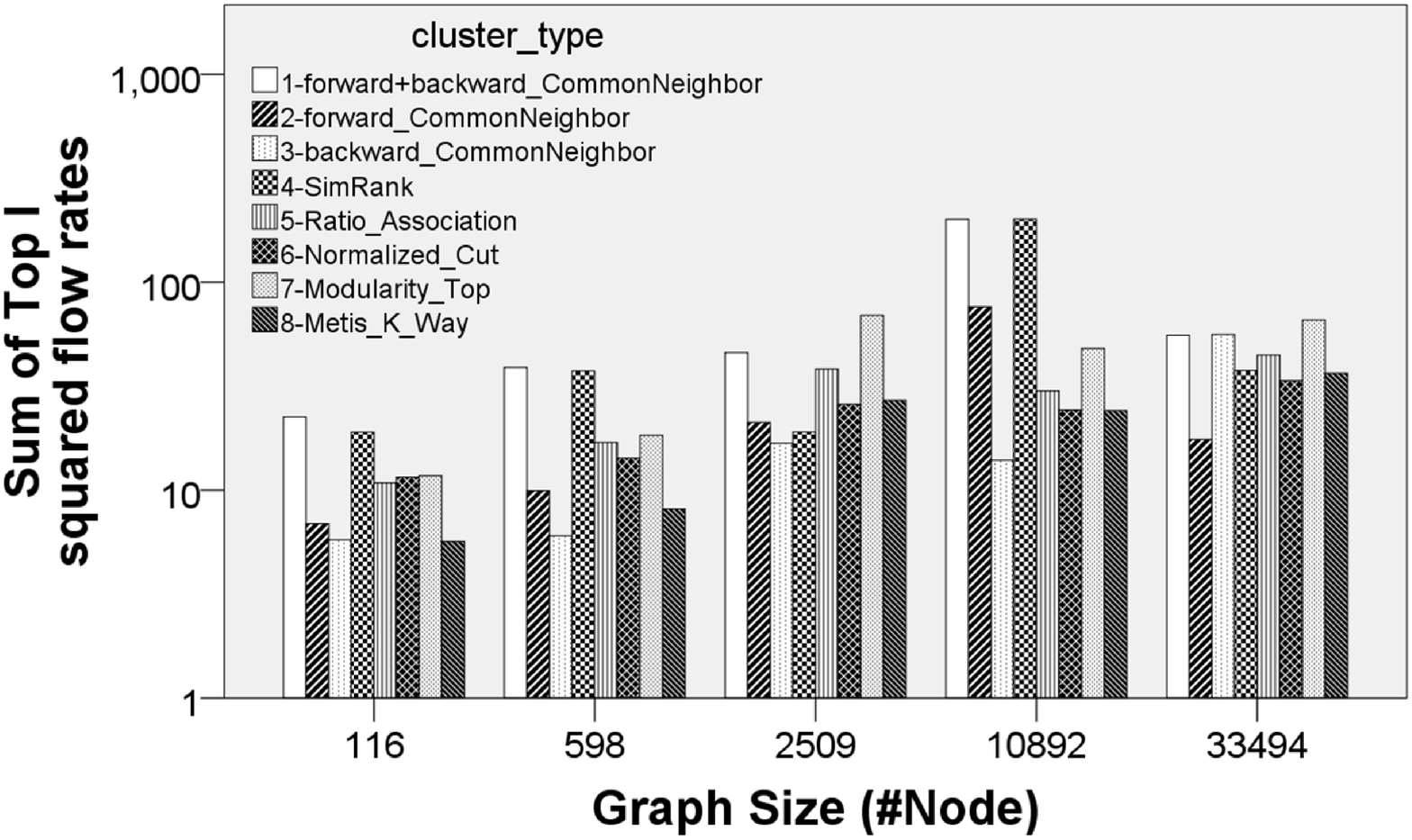}}
\hspace{0.1 in}
\subfigure[$k$ = 10, $l$ = 20]{\includegraphics[height=1.8 in]{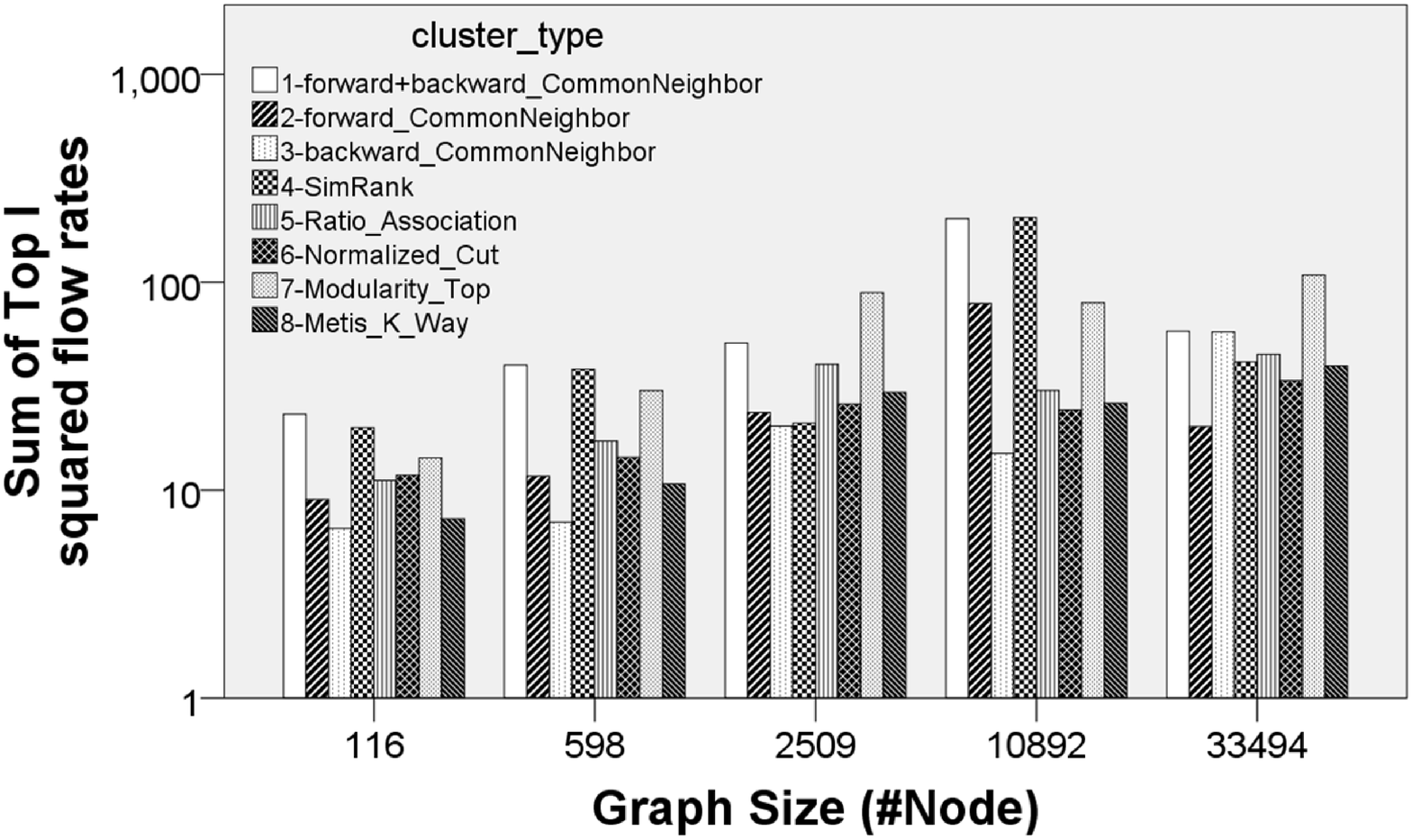}}\\
\vspace{-0.05 in}
\subfigure[$k$ = 20, $l$ = 20]{\includegraphics[height=1.8 in]{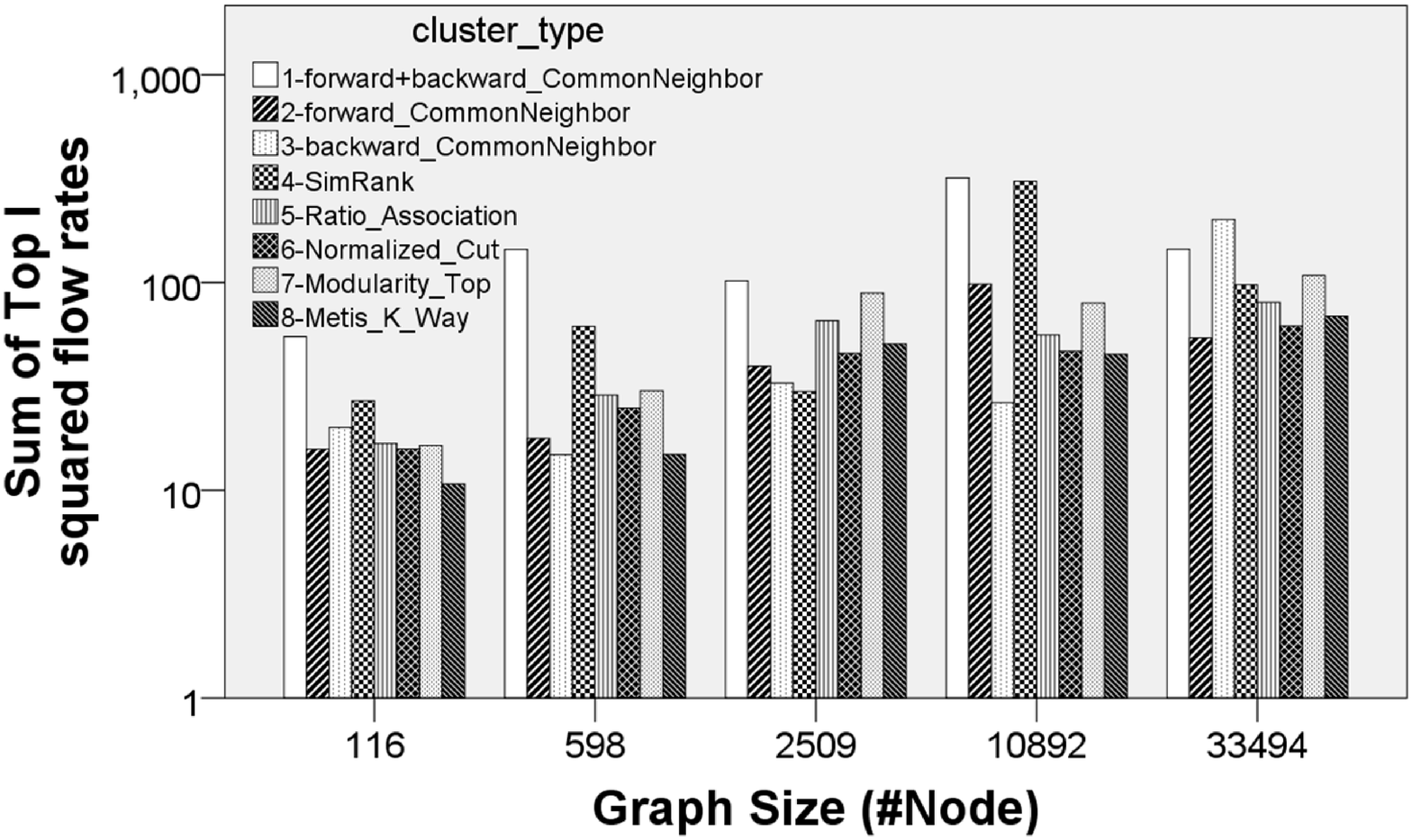}}
\hspace{0.1 in}
\subfigure[$k$ = 40, $l$ = 40]{\includegraphics[height=1.8 in]{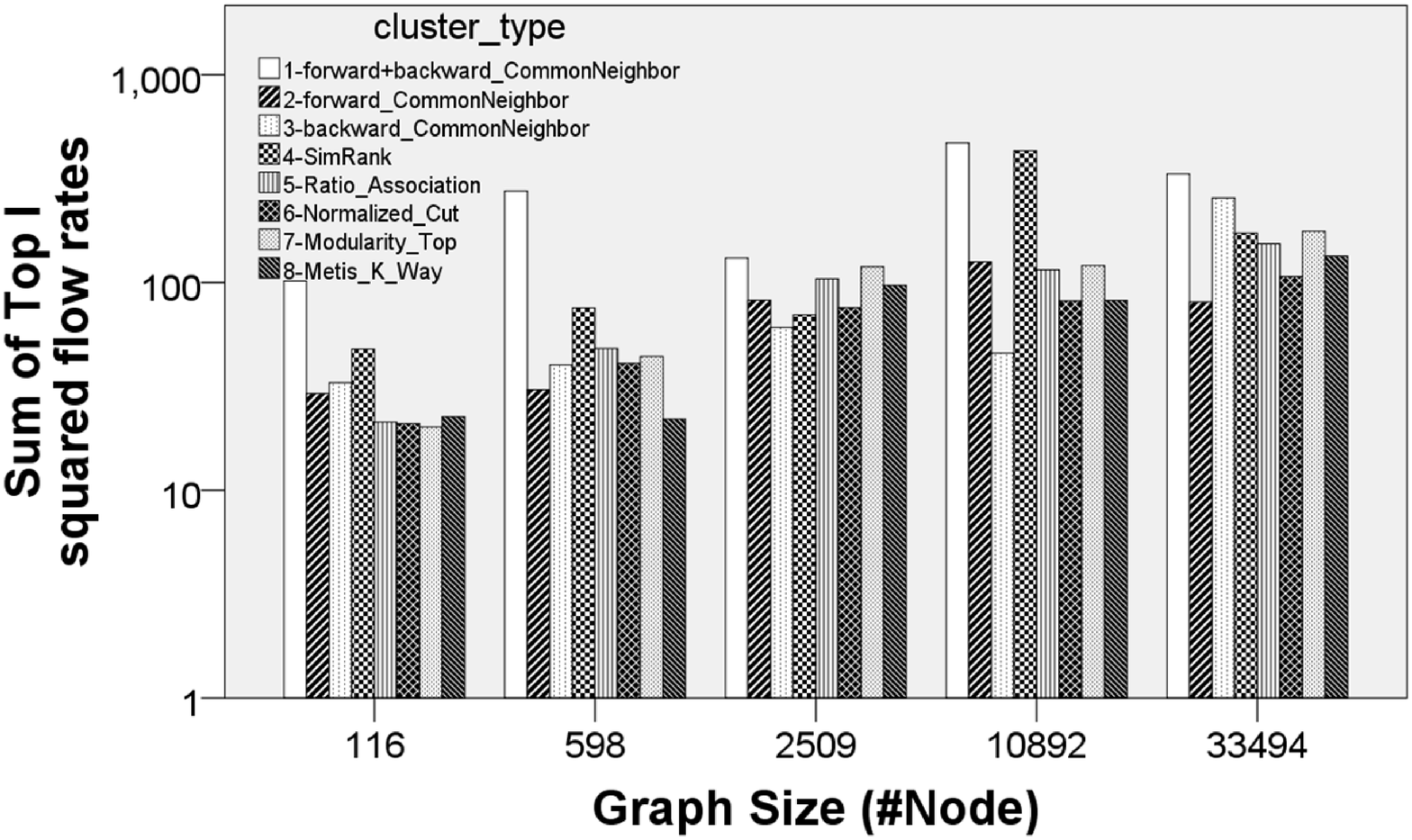}}
\vspace{-0.1 in}
\caption{The performance in maximizing the IGS objective on five sample graphs. The flow rate is summed from the top $l$ flows between all the $k$ clusters. %The weight matrix in each case is computed by RWR with a restart probability of 0.8.
} \label{fig:IGSPerformance}
\vspace{-0.15 in}
\end{figure*}

\bsec{Evaluation}{Eva}

In this section, we evaluate the proposed IGS framework and the CommonNeighbor algorithms by comparing with alternative graph summarization methods. Nine approaches are considered: three using {\em CommonNeighbor} algorithms to compute the similarity matrix for SymNMF (i.e. forward+backward, forward, and backward settings), one using {\em SimRank} algorithm \cite{SimRank} to compute the similarity matrix for SymNMF, the classical graph clustering algorithm with {\em Ratio Association} and {\em Normalized Cut} objectives \cite{NormalizedCut}, agglomerative {\em Modularity}-based graph clustering \cite{Newman:FastAlgorithm}, {\em Metis} K-way graph partition \cite{Metis} and the Minimal Description Length ({\em MDL}) based graph summarization \cite{Sigmod08boundederror}. Note that Ratio Association and Normalized Cut are implemented using their equivalent similarity matrix computation for SymNMF \cite{SymNMF2}. Metis partition is implemented by official open source software package \cite{MetisSoftware}. Modularity clustering is executed agglomeratively until all clusters stop merging at the top level or the number of clusters reaches $k$. For MDL, we implement the greedy algorithm in \cite{Sigmod08boundederror}. The MDL algorithm can not specify the number of clusters, in fact, it generates 4,937 clusters on one medium-sized influence graph. To ensure fair comparison (a larger number of clusters will lead to a much higher overall flow rate), we exclude MDL from numeric comparisons, but present its visual summarization results.

All the experiments are conducted on the same Linux server with two 8-core 2.9GHz Intel Xeon E5-2690 CPU and 384GB of memory. All the LAPACK and ARPACK libs are compiled locally to provide machine-optimized performance. Note that the modularity and Metis implementations are using native-version software package, not guaranteed to be optimized for multi-threading. The raw experiment data are paper citation graphs collected from ArnetMiner \cite{ArnetMiner}. The influence graphs are obtained by reversing the citation links.

\begin{table}
\centering
\caption{Citation graphs used in the experiment.}
\label{tab:CitationGraphs}
\small
\begin{tabular}
{| p{4cm} | p{1.8cm} | c | c | } \hline
Source paper title & Venue/Year & Node & Link \\ \hline \hline
Analysis of a hybrid cutoff priority scheme ... & Wireless Networks 1998 & 116 & 148\\ \hline
Manifold-ranking based image retrieval & ACM Multimedia 2004 & 598 & 895\\ \hline
Stochastic High-Level Petri Nets and Applications & IEEE TC 1988 & 2509 & 5256\\ \hline
Mining Frequent Patterns without Candidate Generation & SIGMOD 2000 & 10892 & 22301\\ \hline
On Power-law Relationships of the Internet Topology & SIGCOMM 1999 & 33494 & 86398\\ \hline
\end{tabular}
\vspace{-0.15 in}
\end{table}

\begin{figure*}[t]
\centering
\includegraphics[width=3.9 in]{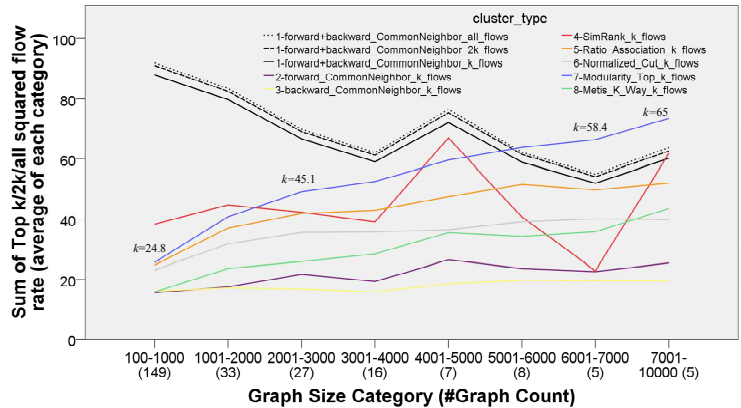}
\vspace{-0.18 in}
\caption{The squared IGS performance of 250 citation graphs with the number of nodes ranging from 100 to 10000. The cluster number is set to $k$ = 20.} \label{fig:IGSPerformance_MultipleGraph}
\vspace{-0.18 in}
\end{figure*}

\bsubsec{Flow Rate Maximization}{Eva-Flow}

We first pick five source papers from the data set to generate maximal influence graphs, as listed in \rtab{CitationGraphs}. These influence graphs are summarized into $k$ clusters, between which the top $l$ flow rates are summed according to the squared IGS objective in \req{SquaredIGS}. \rfig{IGSPerformance}(a)$\sim$(d) present the comparisons among eight summarization methods on the numeric objective function.

The initial result in \rfig{IGSPerformance}(a) with a minimal graph summarization ($k=10, l=10$) suggests that among three CommonNeighbor algorithms, the bidirectional setting almost always achieves the best performance in maximizing the IGS objective (at least $>100\%$ gain \footnote{Percentage of performance gain (drop) by $\frac{new\_number - original\_number}{original\_number} \times 100\%$, the same below.}), except on the largest graph (\#Node=33,494), the backward CommonNeighbor obtains a tiny advantage (1\%). Further, comparing the bidirectional CommonNeighbor to traditional graph summarization methods, CommonNeighbor achieves much better performance than Ratio Association, Normalized Cut and Metis (at least $>20\%$, in average $>100\%$). In some cases, the performance of CommonNeighbor is matched by SimRank ($<10\%$ gain) or outperformed by Modularity.

When we double the number of flows ($k=10, l=20$) in \rfig{IGSPerformance}(b), the sum of flow rates does not increase much on all algorithms (in average $<15\%$) and the overall comparative patterns stay unchanged. This shows that the top $k$ flows already capture most of the flow rates on the graph summarization. We then increase the number of clusters ($k=20, l=20$; $k=40, l=40$). The results in \rfig{IGSPerformance}(c)(d) reveal that the objection function increases much as the number of clusters increases (at least $>30\%$, in average $>90\%$, comparing \rfig{IGSPerformance}(c) with \rfig{IGSPerformance}(b)), except for Modularity, which remains unchanged because their number of clusters are already larger than $k$ and kept stable. For example, the sample graph with 33,494 nodes stops at 71 clusters in the top modularity level. On the comparative pattern, bidirectional CommonNeighbor regains performance advantage over SimRank and Modularity under a large number of clusters.

During the experiment, we have executed each algorithm case three times and report their average performance. However, the results in \rfig{IGSPerformance} still show some randomness due to the nature of iterative NMF solver. To obtain more accurate result, we carefully sample 250 well-cited source papers published in KDD and ICDM from the ArnetMiner data set. The size of their maximal influence graphs are within the range of 100$\sim$10,000 nodes. On each graph, similar experiments are conducted as above given a setting of $k=20$. Finally in \rfig{IGSPerformance_MultipleGraph}, 250 graphs are categorized into 8 bins according to their size. The average performance in each bin are reported for comparison. Results on the larger data set demonstrate the same pattern with the five sample graphs. Bidirectional CommonNeighbor in most cases are the best, except for Modularity, which becomes better as the number of nodes increases beyond 5,000. As mentioned, this is because the Modularity algorithm generates more clusters than the initial setting of $k=20$. As indicated by the labels above the Modularity performance (blue line), the number of clusters increases from 24.8 in the first category to 65 among the largest graphs. Increasing the number of flows for CommonNeighbor does not optimize the objective function much.

\bsubsec{Visualization}{Eva-Vis}

%enhance with topic keyword

\begin{figure*}
\centering
\subfigure[CommonNeighbor (proposed)]{\includegraphics[width=2.1 in]{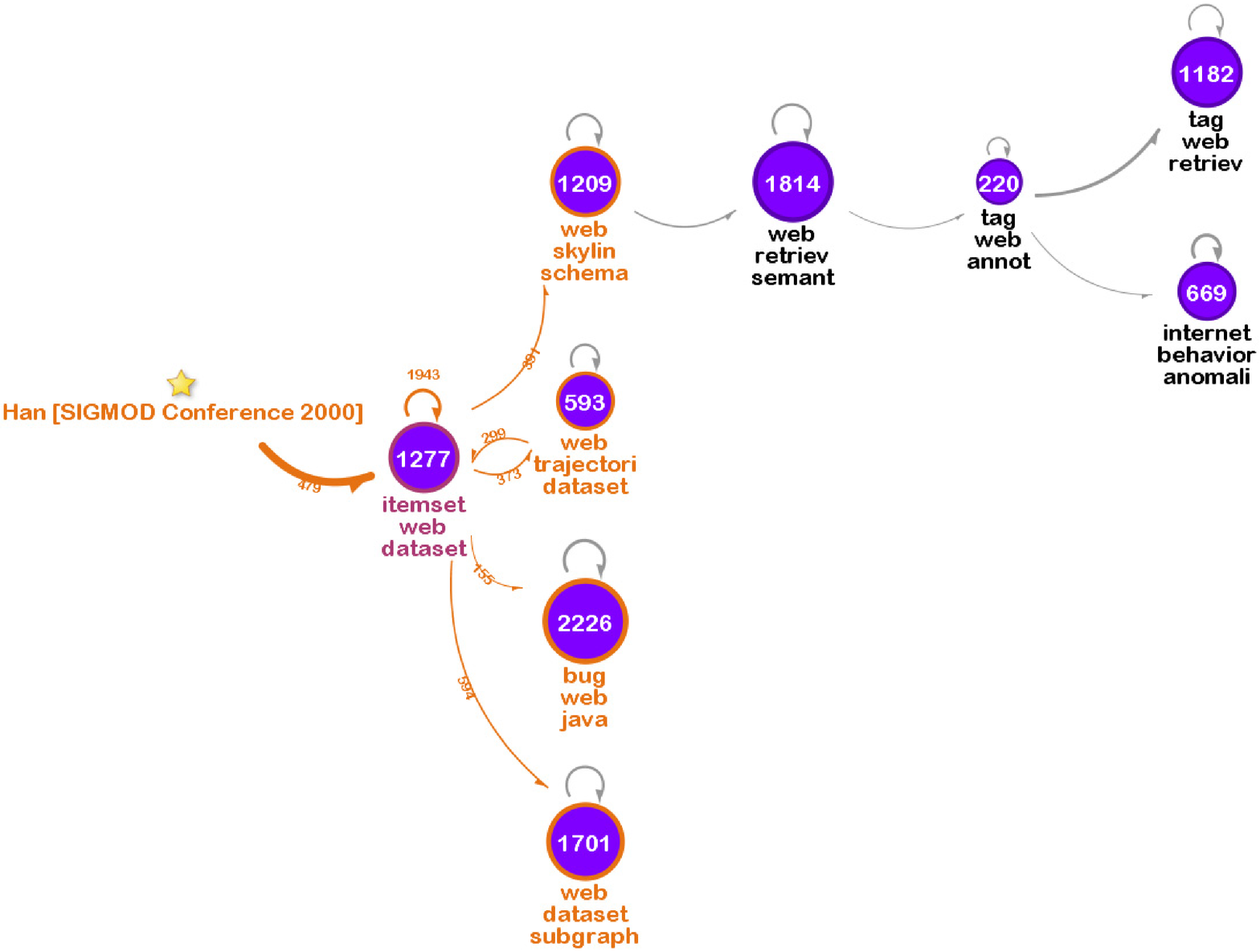}}
\hspace{0.15 in}
\subfigure[SimRank]{\includegraphics[width=2.17 in]{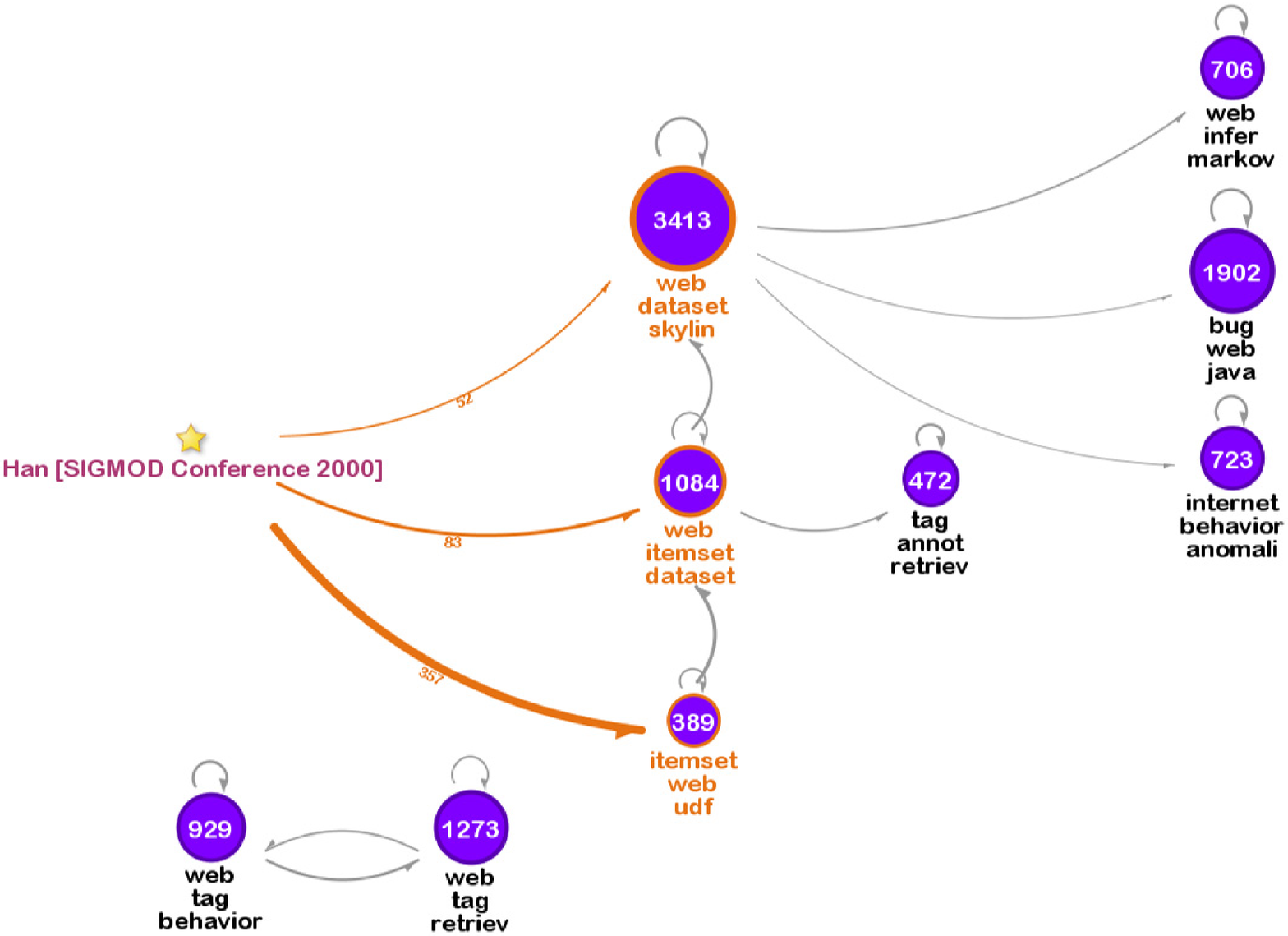}}
\hspace{0.15 in}
\subfigure[Ratio Association]{\includegraphics[width=2.15 in]{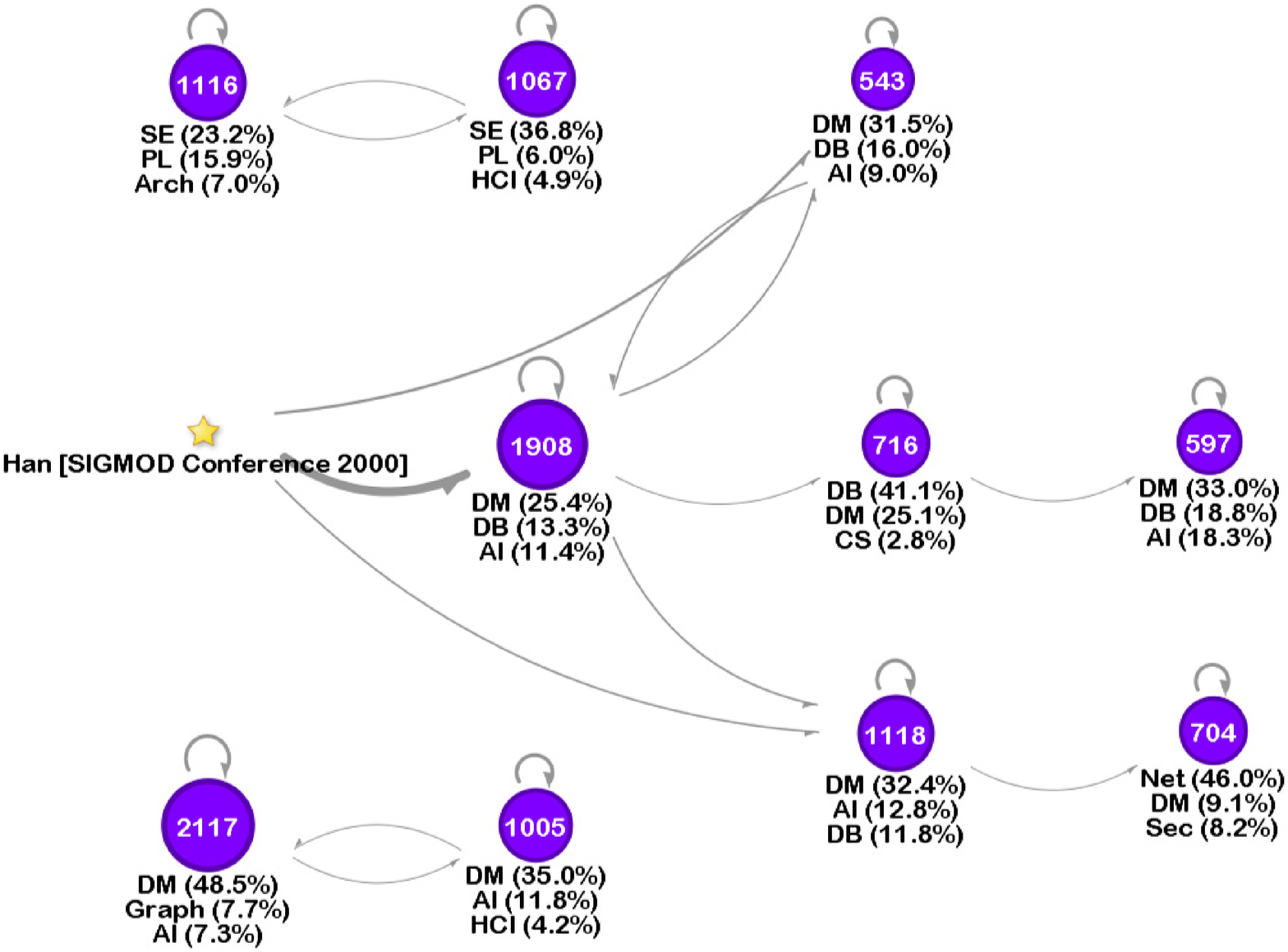}}\\
\subfigure[Normalized Cut]{\includegraphics[width=1.3 in]{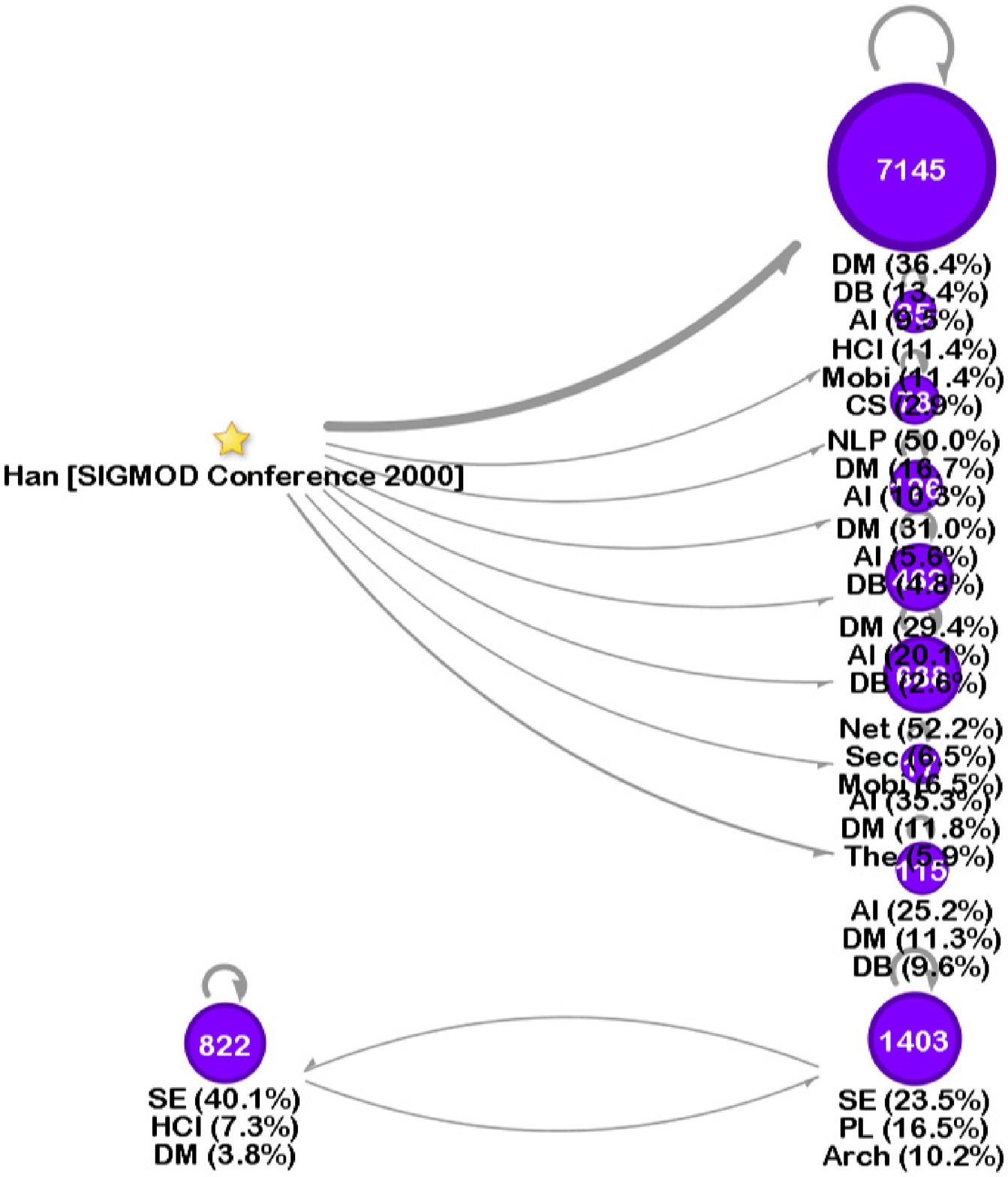}}
\hspace{0.1 in}
\subfigure[Modularity]{\includegraphics[width=1.7 in]{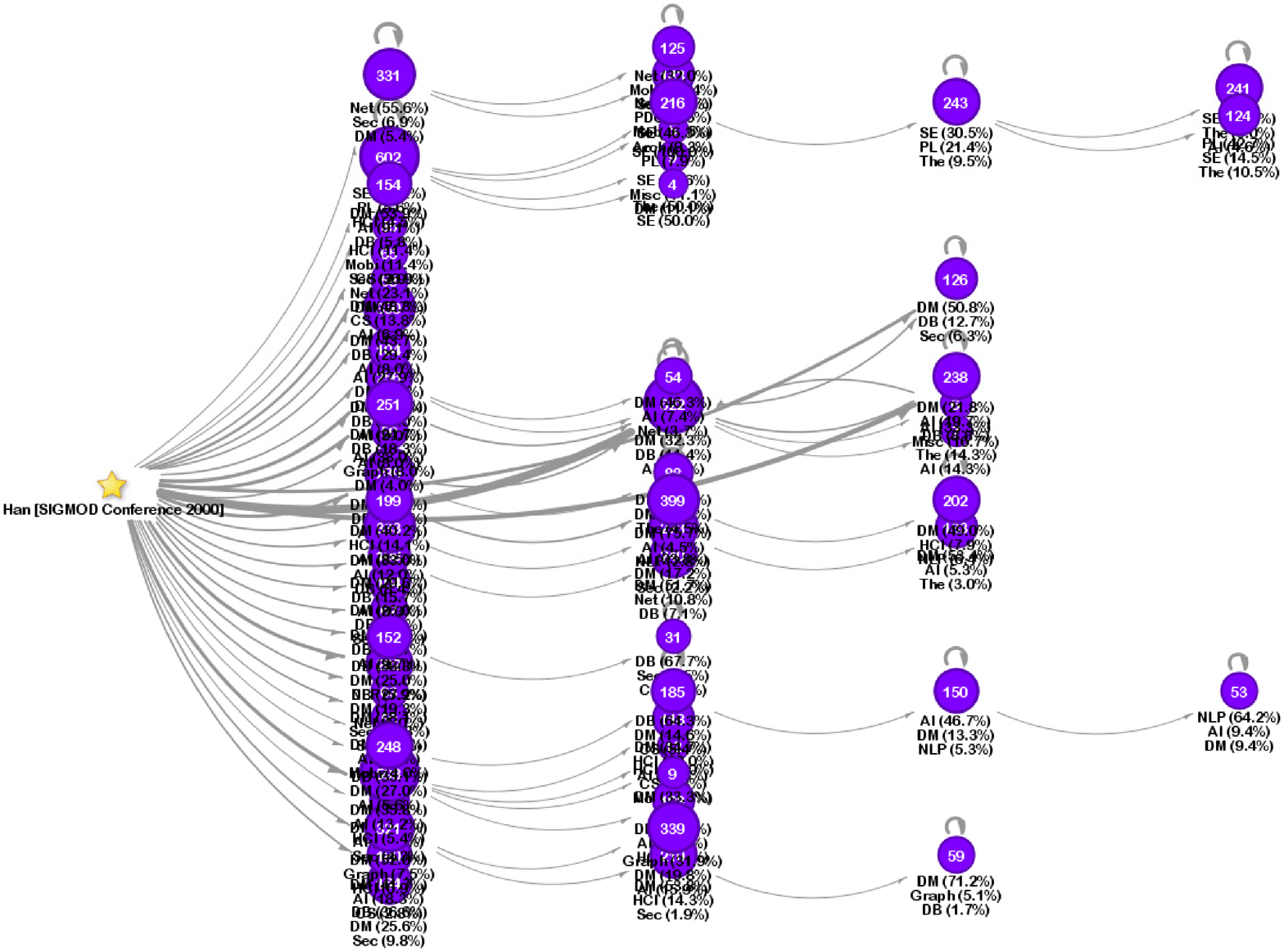}}
\hspace{0.15 in}
\subfigure[Metis K-way]{\includegraphics[width=1.6 in]{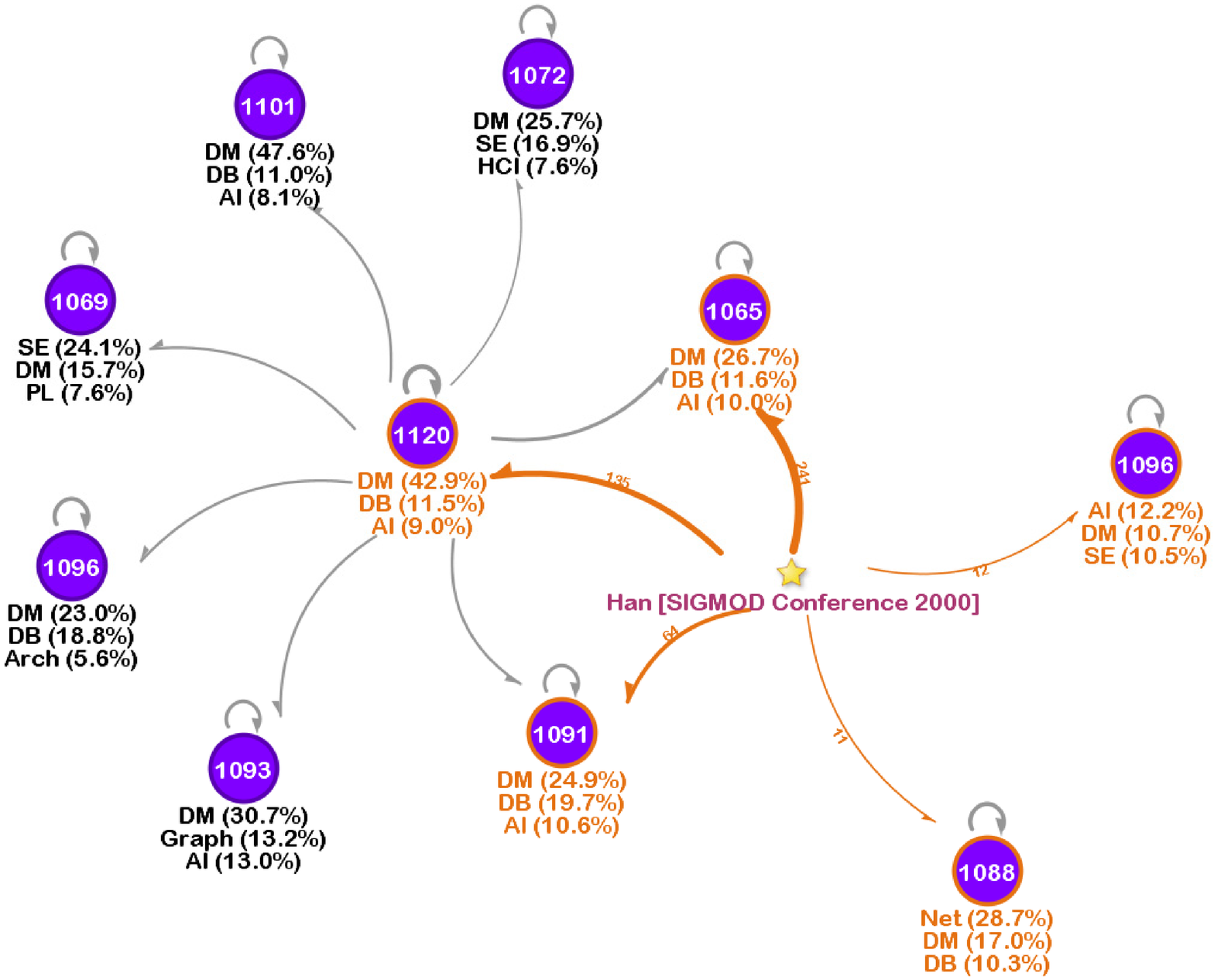}}
\hspace{0.15 in}
\subfigure[MDL]{\includegraphics[width=1.7 in]{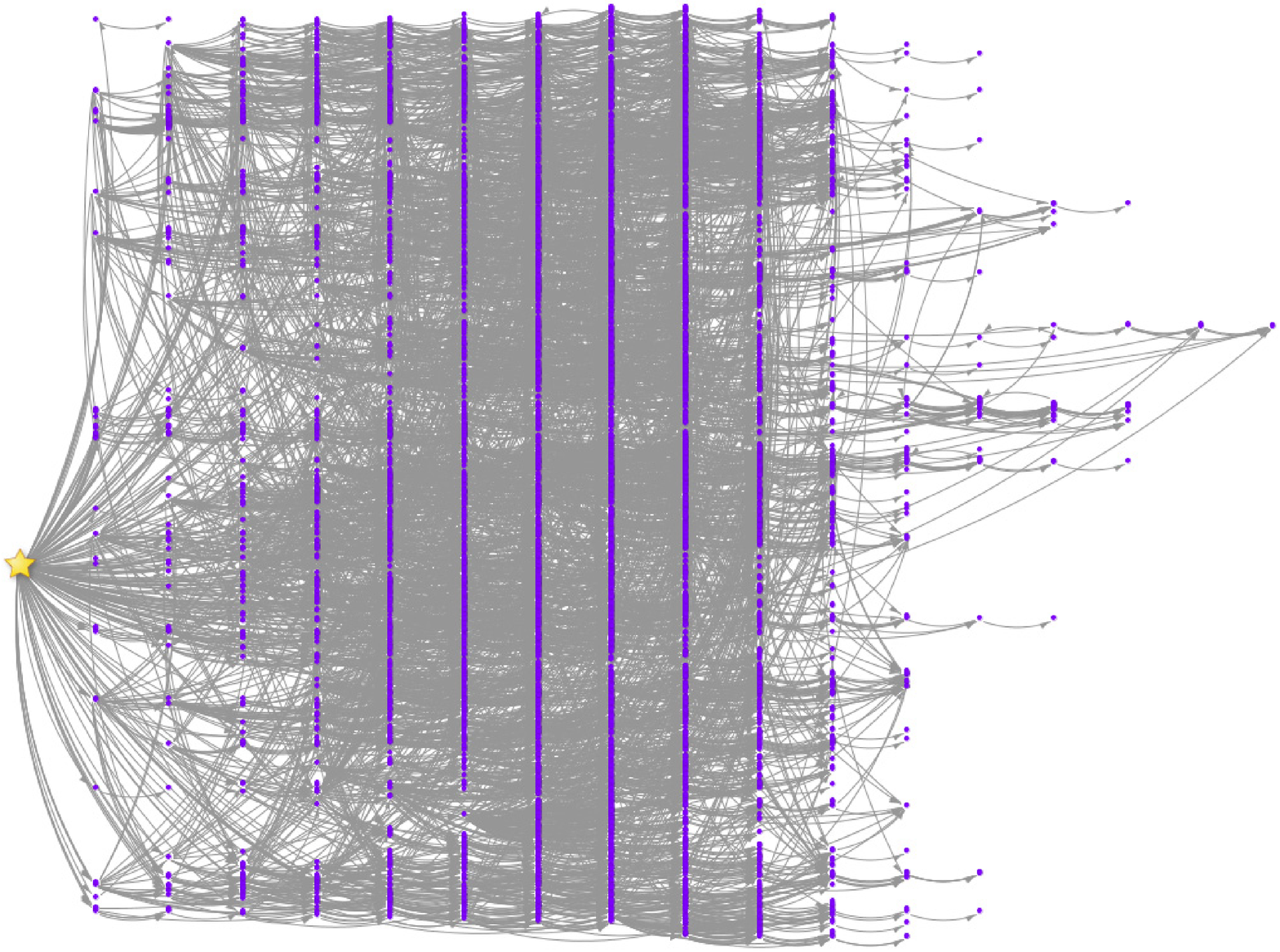}}\\
\vspace{-0.1 in}
\caption{Influence graph summarization results on [Han SIGMOD'2000] by different methods ($k$ = 10, $l$ = 20). Node label gives the number of papers in each cluster and their content summary by either title+abstract keywords in (a),(b) or the top 3 research fields in (c)$\sim$(f). Link thickness indicates the normalized flow rate. Some part of the graph is highlighted to show the number of citations as edge labels. Note that the modularity algorithm stops at 62 clusters and can not merge any further. MDL produces 4,937 clusters, leaving a half of the visual complexity from the input graph.} \label{fig:SIGMODPaperExample}
\vspace{-0.15 in}
\end{figure*}

%Empirical results
We evaluate the effectiveness of summarization methods also by comparing their visualization results: whether they produce a clean influence graph summarization with little visual clutter and whether the results are meaningful for users with domain knowledge. We first pick the famous frequent pattern mining paper by Prof. Jiawei Han et al. as the source to generate the maximal influence graph. Then we execute seven typical summarization methods and depict their results in \rfig{SIGMODPaperExample}(a)$\sim$(g). At the first glance, the proposed bidirectional CommonNeighor method generates a connected tree-like influence graph summarization without edge crossing (\rfig{SIGMODPaperExample}(a)). Compared to that, SimRank gets a similar visual form (\rfig{SIGMODPaperExample}(b)) due to the comparable objective function result, but the generated graph is not connected. The Metis result is also clean (\rfig{SIGMODPaperExample}(f)), but all the clusters have a similar number of nodes, making the summarized graph impractical for usage. Ratio Association and Normalized Cut look inferior due to the poor graph connectivity (\rfig{SIGMODPaperExample}(c)) and the flat influence hierarchy (\rfig{SIGMODPaperExample}(d)). Modularity and MDL are the worst because of the visual clutter generated from the large number of clusters remained in the summarization (\rfig{SIGMODPaperExample}(e)(g)).

Taking a closer look at the visual summarizations, we find that by CommonNeighbor, most flows represent at least 300 citation links. While by SimRank, the critical flows linking the source node are fragmented, two of which only include 52 and 83 citations. The same deficiency is found in the result by Metis, where two highlighted flows only have 11 and 12 citations. We also invite a senior researcher from the database and data mining community to evaluate the summarization result. With our interactive tool, she can switch between the title+abstract summary and the research field summary. She can also access paper details in each node cluster with a sorted list by citation count. She mainly compares the visual summarization by CommonNeighbor and SimRank. In this case, she prefers the result by CommonNeighbor in \rfig{SIGMODPaperExample}(a) because the influence evolutions make more sense: the initial paper quickly raises much attention on pattern mining research such as itemset and association rule mining, then the thread splits into four streams on general data management research (such as web and uncertainty skyline analysis), trajectory analysis, subgraph analysis and application in software engineering (e.g. bug analysis). The thread of web data analysis gradually moves to web retrieval and finally leads to tag analysis and anomaly behavior detection. Compared with CommonNeighbor, SimRank creates some false links, e.g. the direct flow from the frequent pattern mining paper to uncertainty data analysis.

Furthermore, we ask our invited users to study the influence of the well-known Internet power-law paper in SIGCOMM'1999. The maximal influence graph is summarized by the bidirectional CommonNeighbor algorithm into \rfig{SIGCOMMPaperExample_20} (in the second page). Note that in this case the influence graph topology is augmented by the ``venue'' field of each paper to group the papers with similar research topic together. From the visual summarization, she learns that the SIGCOMM paper directly influences the research on Internet topology and simulation. Next, over the Internet topology topics, the P2P research becomes popular and after that the web-related research and XML. The most recent hot topic in this thread appears to be sensor network which corresponds well to his domain knowledge.

\begin{figure}
\centering
\includegraphics[width=3.2 in]{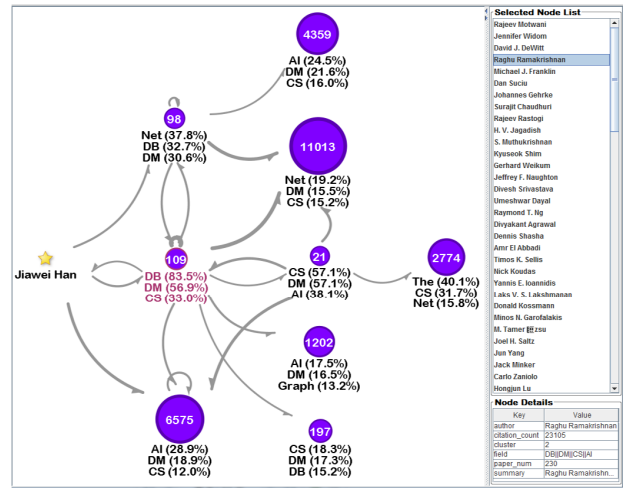}
\vspace{-0.1 in}
\caption{The summarization of Prof. Jiawei Han's influence graph by bidirectional CommonNeighbor ($k = 10$).} \label{fig:HanAuthorExample}
\vspace{-0.2 in}
\end{figure}

Our framework can also visualize one author's influence by summarizing the author influence graph. This graph is generated by adding one influence link between two authors for each citation between their papers. The maximal author influence graph is then computed from a source author by traversing the influence graph. As an example, we select Prof. Jiawei Han as the source author, and collect the influenced authors within 2 hops. To limit the size of the influence graph, we only keep productive authors (i.e. $\geq 30$ paper publications) which gives a graph of 26,349 author nodes. The summarization result applying bidirectional CommonNeighbor algorithm ($k=10$) is shown in \rfig{HanAuthorExample}. Our invited user acknowledges the validity of the result: Prof. Han has influenced multiple fields with his research, mainly data mining (DM), database (DB), AI and networking (Net). On his contribution to DB and DM fields, the influence is bidirectional, i.e. he is also heavily influenced by the researchers there, as indicated by the group of 109 authors in the picture (e.g. Raghu Ramakrishnan). The most directly influenced field by the number of authors are AI and DM, as indicated by the group of 6,575 authors. The most indirectly influenced field are Net and DM, as indicated by the group of 11,013 authors. Through the bridging of a group of 21 authors (e.g. Rakesh Agrawal), he also impacts the Theory (The) research, represented by the group of 2,774 authors.

\bsubsec{Scalability}{Eva-Case}

\iffalse
\begin{figure}
\centering
\includegraphics[width=3.3 in]{pic/TotalTime.eps}
\vspace{-0.15 in}
\caption{The total computation time of summarization methods, $k$ = 20.} \label{fig:OverallTimePerformance}
\vspace{-0.1 in}
\end{figure}
\fi

%time performance

\begin{figure}
\centering
\subfigure[Total computation time (second) by graph size (\#node)]{\includegraphics[width=3.3 in]{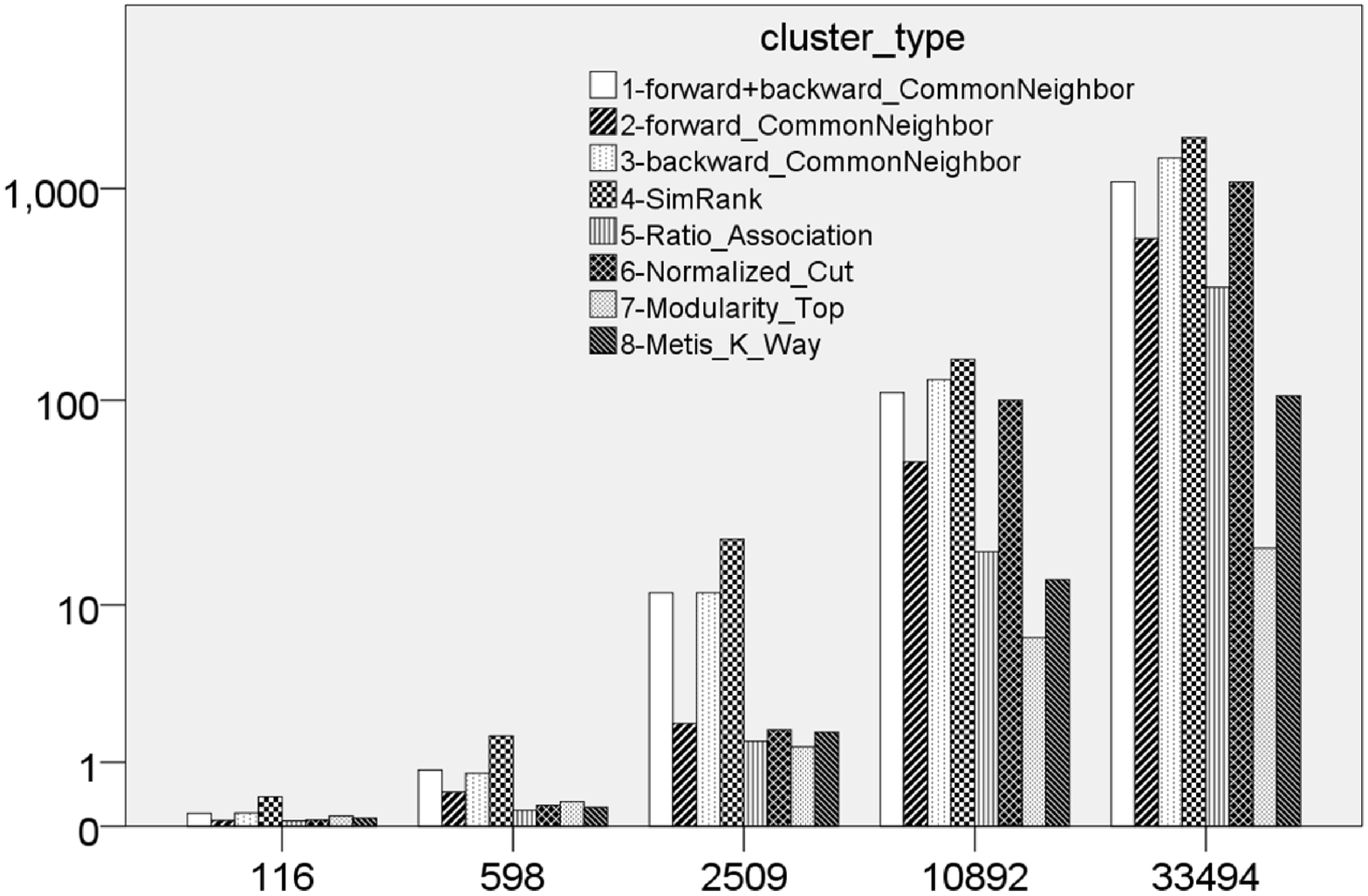}}\\
\vspace{-0.1 in}
\subfigure[Similarity Matrix]{\includegraphics[height=1.05 in]{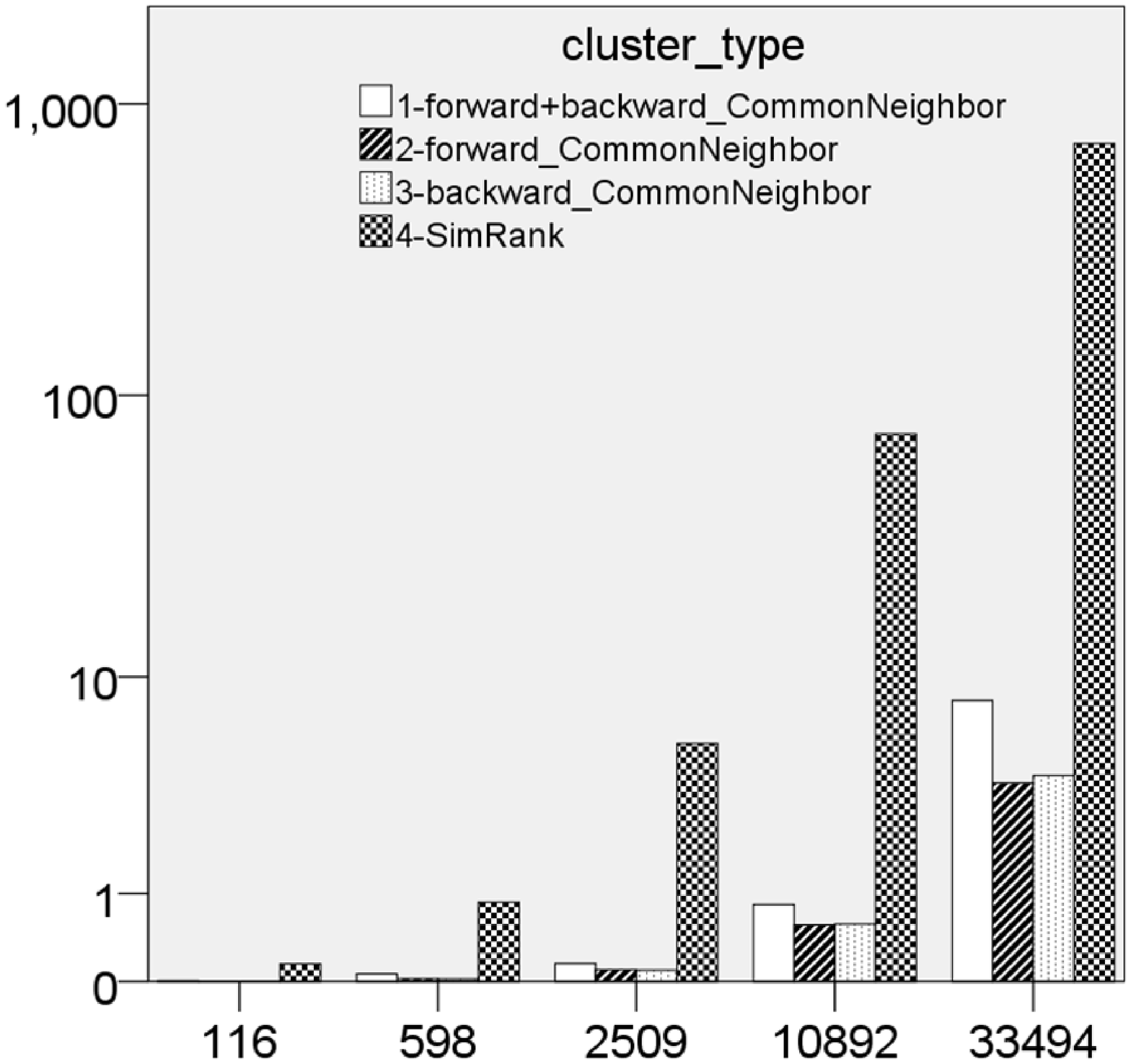}}
\subfigure[Eigen Initialization]{\includegraphics[height=1.05 in]{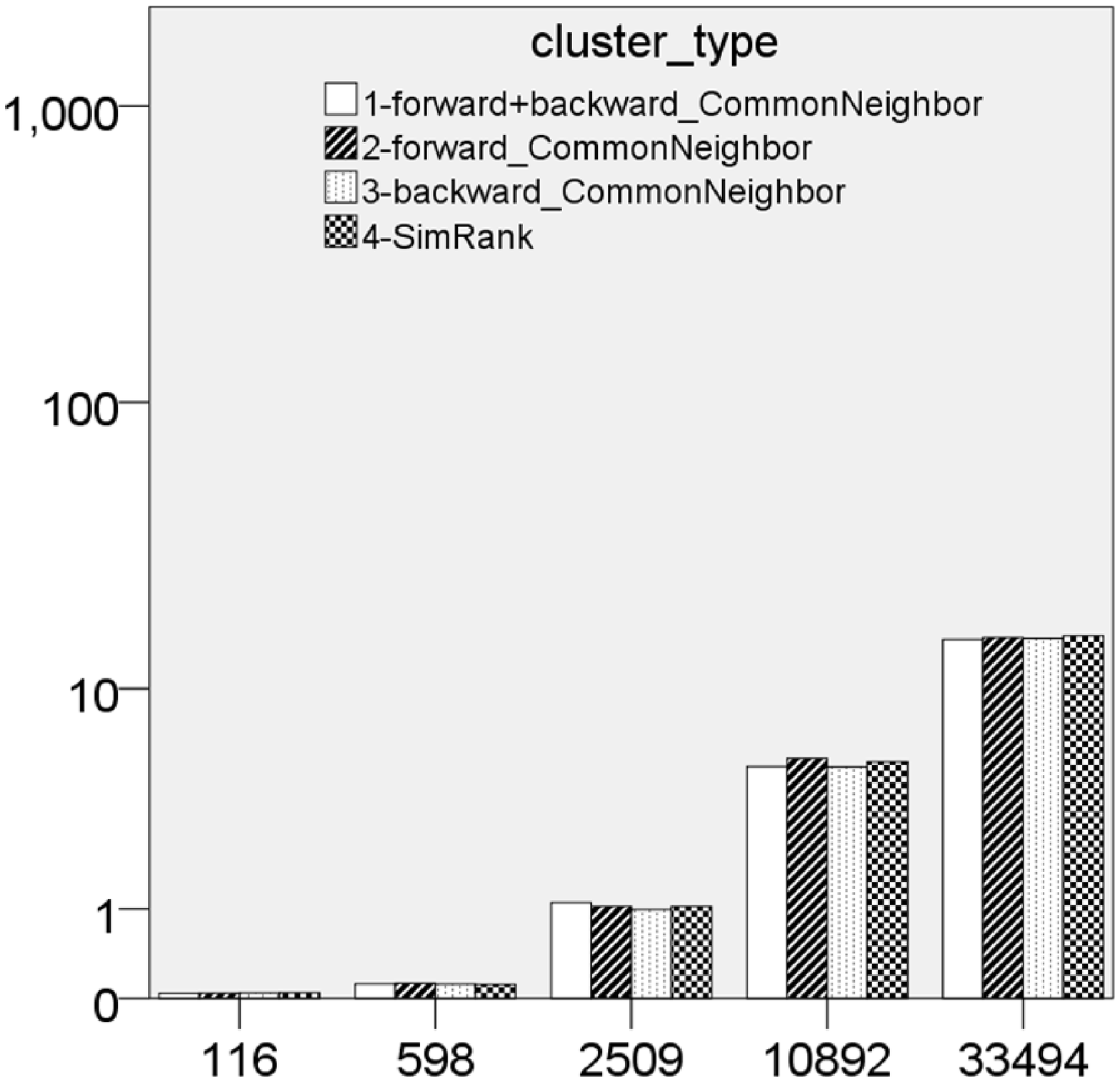}}
\subfigure[SymNMF]{\includegraphics[height=1.05 in]{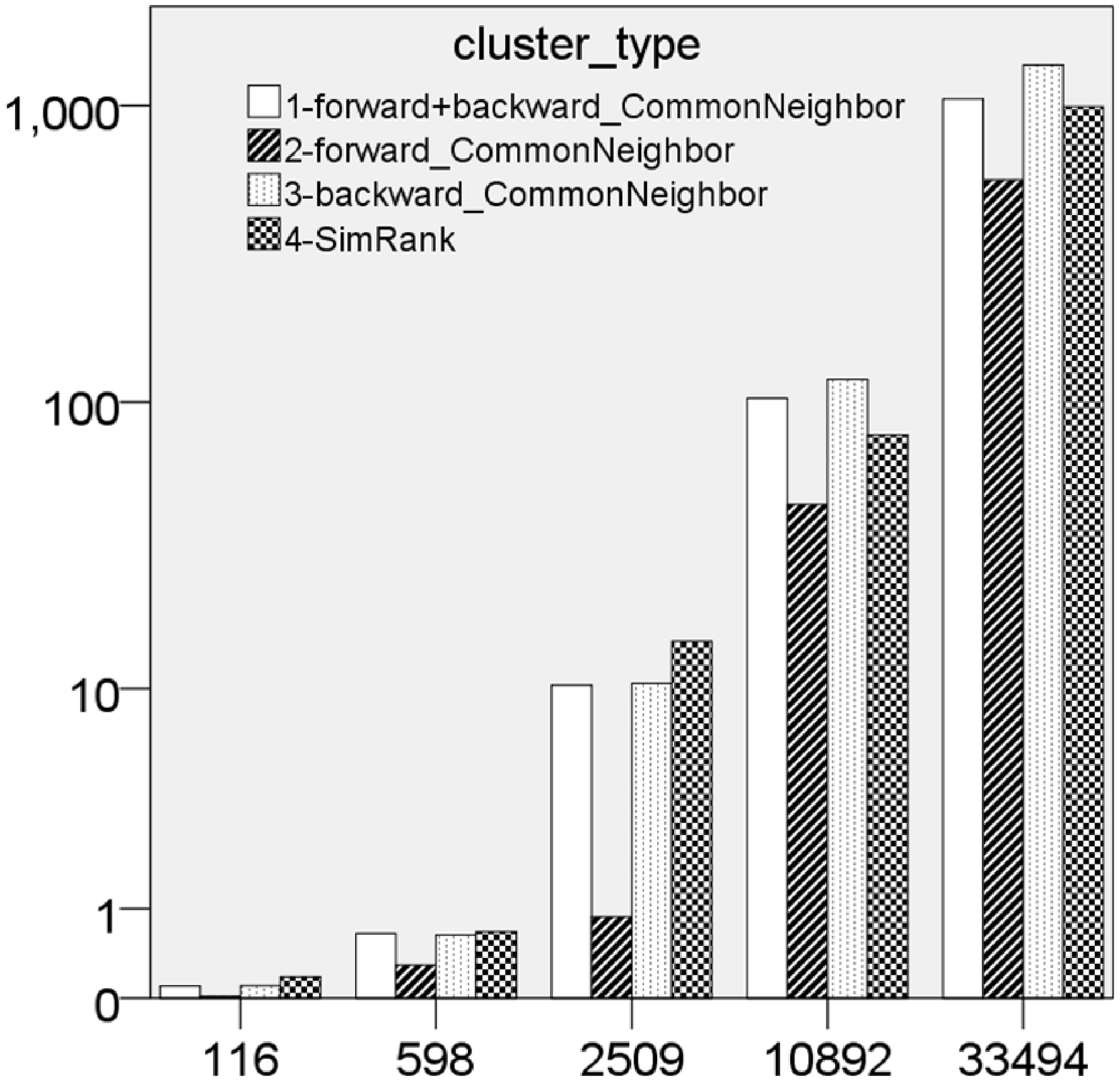}}
\vspace{-0.1 in}
\caption{The time cost of different summarization methods, $k$ = 20: (a) Total time; (b)$\sim$(d) Split time of four algorithms in our framework. The similarity matrix computation and SymNMF iteration dominate the cost.} \label{fig:TimePerformance}
\vspace{-0.15 in}
\end{figure}

The overall computation time for different summarization methods is illustrated in \rfig{TimePerformance}(a). Our proposed algorithms are more costly than efficient modularity clustering algorithm ($O(nlog(n))$ with small constant) and Metis k-way graph partition ($O(n+m)$). However, the best of our methods can summarize a 10,000-node maximal influence graph in 100 seconds, and the overall time complexity is only slightly above linear. Note that $n$ here denotes the size of the maximal influence graph, which is much smaller than the size of the original graph. Most citation graphs from a single paper are no larger than the magnitude of 10,000 nodes, while the entire data set has millions of papers.

Within our framework, the SimRank algorithm requires the longest computation time. To explain this, we have looked at the split time at three key steps, as shown in \rfig{TimePerformance}. The eigenvalue decomposition (\rfig{TimePerformance}(c), only top $k$ eigenvectors are computed) are quite fast due to the sparsity of the influence graph matrix (\rtab{CitationGraphs}). On similarity matrix computation (\rfig{TimePerformance}(b)), SimRank is slow because in worst case it needs to compute an all-to-all similarity matrix ($O(n^2d^2)$), though we have optimized it to only compute within a 4-hop range. In contrast, CommonNeighbor is much faster on similarity computation, through the multi-threaded routine on sparse matrix multiplication. Finally, SymNMF (\rfig{TimePerformance}(d)) is the most costly step. In each iteration, there are a few sparse matrix-matrix multiplication computations.

Compared with the time complexity, the space requirement of our framework is less stringent. The similarity matrix computation and iterative SymNMF each needs to store a dense matrix at most, giving a space complexity of $O(n^2)$ with small constant. The eigenvalue decomposition by dsyevx routine in LAPACK only needs $O(n)$ space with a relatively large constant. Recall that $n$ is the number of nodes in the maximal influence graph and can be hundreds of times smaller than the original input graph. %In our experiment runtime, a 40GB memory is enough to summarize the largest graph with 33,494 nodes.

\bsubsec{Summary and Discussions}{Eva-Discuss}

First, our experiment results demonstrate that the summarization methods specifying the number of clusters provide compact influence graph summarizations. In contrast, typical graph compression and summarization methods such as MDL and Modularity can lead to huge visual clutters that make it hard for user to interpret. Within the $k$-cluster methods, applying bidirectional CommonNeighbor algorithm in our framework is shown to be the best in maximizing the IGS objective, constantly superior than traditional graph partition and clustering algorithms, such as Ratio Association, Normalized Cut and Metis. In a few cases, plugging SimRank into our framework can achieve comparable performance. In fact, SimRank has very close tie to our method. CommonNeighbor considers the similarity of two nodes in one hop beyond, while SimRank computes their similarity in an infinite hop (pruned to four hops in this work). Our results show that, though close to, SimRank is not better than CommonNeighbor in maximizing the IGS objective, but also it suffers from a higher computational complexity of $O(n^2d^2)$.
%However, it is still unclear how different settings of SimRank will perform (we implement bidirectional SimRank) and whether SimRank can be better in theory.

Second, we note that the parameter $k$ and $l$ in the summarization can be critical for both the objective and user performance. As $k$ becomes large, for example from ($k=20, l=40$) to ($k=40, l=40$) in the cases of \rtab{CitationGraphs}, the IGS objective increases 56\% in average while the visual complexity doubles. We recommend to set $k \leq 20$ on this trade-off, because at a size larger than 20, the node-link graph visualization may not be a good choice for many graph visual analysis tasks \cite{MatrixUX04}. On the choice of $l$, we find that the optimization of IGS objective is not significant after $l \geq 2k$, therefore $l = 2k$ can be an appropriate setting.

Last, we target on the academic data sets in this work. It seems straightforward to apply our framework also to social influence graphs on Twitter and Facebook. However, we caution that the basic retweeting influence graph can be summarized clearly by MDL or structural equivalence \cite{LWEquiv} based node grouping, because such graphs are all standard trees with large structural redundancy. %In future, we plan to study more sophisticated author influence graphs on social media by integrating one's retweeting trees together.

\iffalse
Overall, we recommend bidirectional CommonNeighbor in practice for its good balance among the visualization result, optimization quality and computational efficiency.
\fi

\bsec{Related Work}{Related}

%We review the related work from three aspects. %This work applies visualization-driven objective to the influence graph summarization problem, hence is related to three research areas.

First, {\em graph summarization}, constructing a smaller abstraction to represent the large graph has been a traditional research topic, e.g. using graph clustering algorithms. These algorithms usually optimize certain association or cut measure during the k-way graph partition. Several measures have been proposed, e.g. ratio association, ratio cut \cite{RatioCut} and normalized cut \cite{NormalizedCut}. The similar problem is also studied in the context of community detection by interdisciplinary researchers \cite{CommunityDetectionSurvey}, in which modularity is one of the most popular quality function to access a community \cite{Newman2004}. However, most of the clustering and community detection methods on graph target at maximizing intra-cluster connections while minimizing inter-cluster connections. This is fairly different from the IGS problem studied here. On the other hand, there are also plenty of works in compressing large graphs for efficient storage and representation. In \cite{Sigmod08boundederror}, MDL-based compression was proposed to present the graph with an aggregated structure and an error correction list. It is proved to be the best summary from the information-theoretic objective. While MDL approach can successfully compress web graphs, on influence graphs which are much sparser (the citation graphs have an average degree of less than 3), it performs similarly to a structural equivalence based grouping \cite{LWEquiv}, leaving huge visual clutters unsettled. Another algorithm, SNAP \cite{Sigmod08Tian}, considers the node attribute on graph, but again is not tailored for the influence graph scenario.

Second, {\em visualization}, over the past few decades, the methodology to draw node-link graphs has reached its maturity. On graphs with less than a few hundred nodes, the planar graph drawing approach \cite{GraphDrawingBook} and the force-directed algorithm \cite{KKLayout} can produce visually pleasant graph layouts in real time, mainly by minimizing edge crossings. On large graphs with a thousand or more nodes, the force-based algorithms can be extended by multilevel coarsening and fast force approximation \cite{HuMultiLevel} and still generate a layout in reasonable time (e.g. less than a minute for million-node graphs). However, on real-world large graphs with small-world nature, including the influence graph discussed here, the resulting graph layout still has numerous edge crossings. This leads to overwhelming visual clutters detrimental to visual data mining tasks. Meanwhile, Shahaf et al. \cite{Cartography13}\cite{Thought12}\cite{Connecting10} studied the similar problem of summarizing large amount of information into user-friendly visual maps. They developed intriguing methods to detect hidden linkage and document clusters from the keyword frequency statistics. On a quite different focus, our method is built on the graph with explicit linkage data while the textual content of each node can be absent or incomplete.

%Some recent literatures try to create graph visualizations over existing summarization results (e.g. graph clustering \cite{ASKGraph}\cite{HiMap09}). They improve the viewing experience on large graphs, but neither do they develop any new graph summarization method, nor are they optimized for influence graph visualization.

% or influence graph analysis?
Third, considerable work has been conducted for studying the effects of {\em social influence}.
For example, Bakshy et al. \cite{Bakshy:12EC} conducted randomized controlled trials to identify the effect of social influence on consumer responses to advertising. Bond et al. \cite{Bond:12Nature} used a randomized controlled trial to verify the social influence on political voting behavior. Tang et al.~\cite{Tang:09KDD} presented a Topical Affinity Propagation (TAP) approach to quantify the topic-level social influence in large networks. Kempe et al.~\cite{Kempe:03} proposed to use a submodular function to formalize the influence maximization problem and develop a greedy algorithm to solve the problem with provable approximation guarantee. Most of these works focus on the existence of social influence or the nature of the information diffusion process and do not consider the summarization problem. Recently, Mehmood et al. proposed CSI \cite{CSI13}, a model that generalizes the classical Independent Cascade model to the community level, built from the cascade-based community detection method \cite{Cascade13}. CSI can produce similar visual forms to our result. However, the CSI model is computed from the probabilistic social influence graph and the information propagation log more engaged to the social influence scenario. In comparison, our method is more focused on the visual summarization of large influence graphs in the objective of maximizing flows. We do not leverage the information propagation model and the associated log data in such scenarios.

%To be specific, while these existing work aim to detect {\em which} nodes are most influential, the proposed work aims to explain/interprete {\em what makes them influential} and {\em how they influence others}. %is orthognoal to these existing

%

\bsec{Conclusions}{Con}

In this paper, we propose the influence graph summarization problem, study its linkage to the existing clustering methods, and present a unified framework to solve it. The framework achieves all the three design objectives, including (1) flow rate maximization that highlights the evolution of influence; (2) a localized visualization from the source node; and (3) easy to incorporate rich information on graph such as node attribute and time. The framework is comprehensive and flexible. We provide both the SymNMF based solution and implementation details. Through comprehensive evaluations with real-world academic citation graphs, we demonstrate that our framework constantly outperforms classical methods, such as graph clustering and compression algorithms, in both quantitative performance and qualitative visual effects.

%\section*{Acknowledgment}



\bibliographystyle{IEEEtran}
\bibliography{arxiv}

\end{document}